# Title: First-Principle Modeling Framework of Boost Converter Dynamics for Precise Energy Conversions in Space

**Authors:** Yifan Wang[1], Wenhua Li[1], Zhenlong Wang[1], Xinrui Zhang[1], Jianfeng Sun[1], Qianfu Xia[1], Zhongtao Gou[1], Jiangang Rong[2], Tao Ye[1]*

**Affiliations:**

[1]Ministry of Education Key Laboratory of Micro/Nano Systems for Aerospace, Key Laboratory of Micro- and Nano-Electro-Mechanical Systems of Shaanxi Province, School of Mechanical Engineering, Northwestern Polytechnical University; Xi'an, 710000, China.

[2]Shanghai Guoyu Zhilian Aerospace Technology Co., Ltd; Shanghai, 201112, China.

*Corresponding author. Email: yetao@nwpu.edu.cn

**Abstract:** Boost converters are essential for modern electrification and intelligent technologies. However, conventional Boost converter models relying on steady-state assumptions fail to accurately predict transient behaviors during input voltage and load fluctuations, which cause significant output voltage overshoots and instability, resulting in failures of electrical systems, thereby restricting their use in space. This study introduces a first-principle modeling framework that derives precise dynamic equations for Boost converters by incorporating non-ideal component coupling. As compared to the most accurate existing Boost converter model, the proposed models reduce steady-state and dynamic-state errors between experimental and simulated output voltages by factors of 11.0 (from 20.9% to 1.9%) and 15.4 (from 77.1% to 5.0%) under input voltage variations, and by factors of 10.2 (from 15.3% to 1.5%) and 35.1 (from 42.1% to 1.2%) under load changes, respectively. Consequently, a reliable Boost converter is accordingly designed and on-orbit deployed for precise energy conversions.

**Main Text:**

In the eras of electrification and intelligence, Boost converters uniquely meet the demand for voltage step-up thus serve as cornerstones of modern electrical technologies (*1*, *2*). As the fundamental components of almost all electrical systems, the efficiency and reliability of Boost converters are critical to key sectors from microelectronic chips to electric vehicles and automation robots, from Internet of Things (IoT) nodes to large-scale renewable energy grids and communication infrastructures (**Fig. 1**) (*3*–*5*).

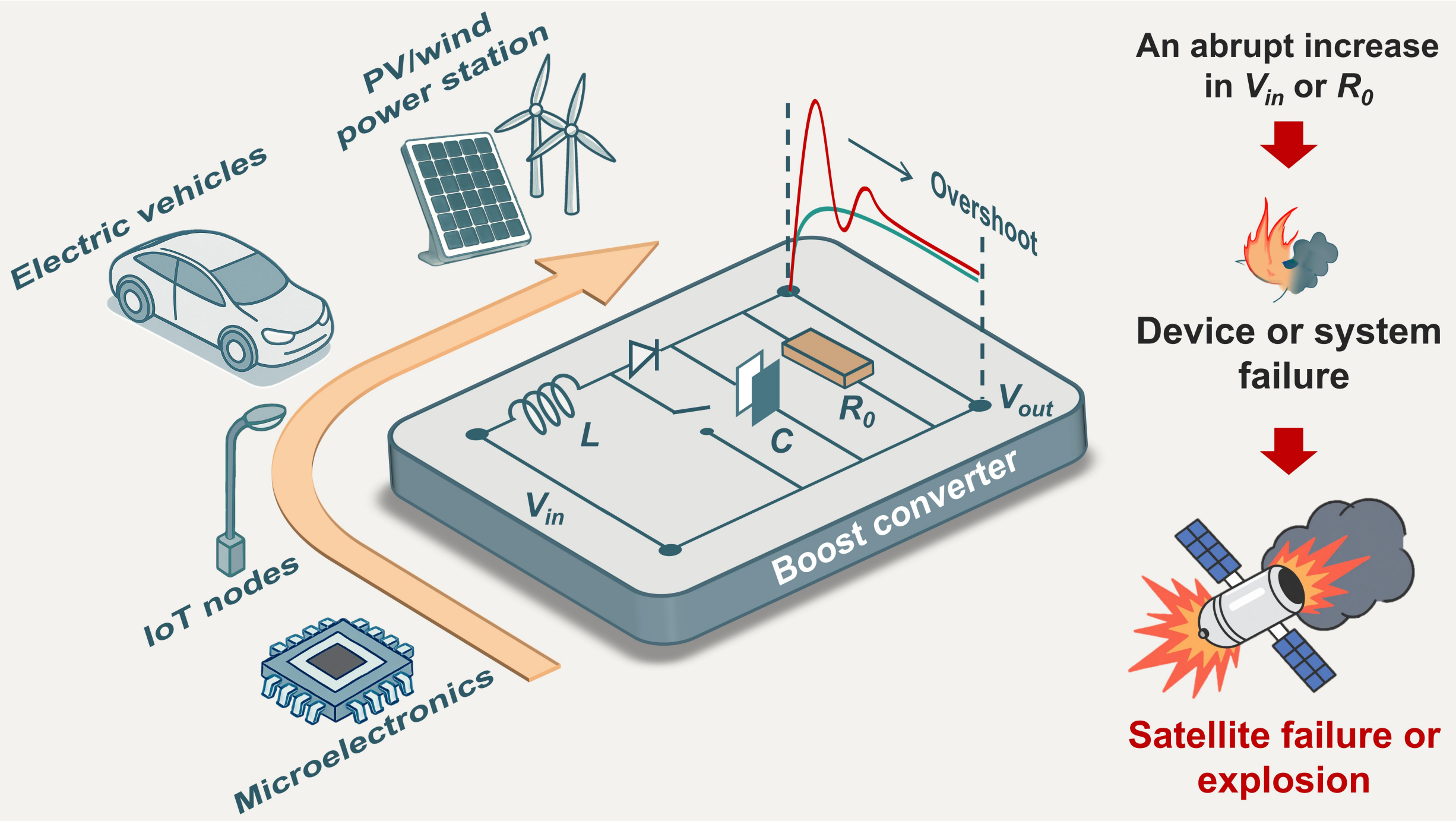

**Fig. 1. Limitations of conventional Boost converter models.** Boost converters are foundational elements of modern electronic systems, underpinning microelectronics, IoT nodes, electric vehicles, and photovoltaic (PV)/wind power stations. Under dynamic operating conditions, the transient output voltage can depart markedly from idealized predictions. Abrupt changes in input voltage or load resistance drive pronounced output voltage overshoots, resulting in component stress and device or system failure.

The performance of Boost converters directly impacts the operational stability, longevity, and energy efficiency of entire electrical systems. Conventional Boost converter models (*6*, *7*), grounded in steady-state assumptions and linear approximations, have been long served as the industry standard. However, these approaches fundamentally failed to capture the intricate dynamics during transient events, such as start-up transients, input voltage fluctuations, or rapid load changes. Thus, the Boost converters designed with conventional models could result in substantial output voltage overshoots (*8*), which might exceed the threshold voltage of the components, resulting in catastrophic failure of electrical systems **(Fig. 1)** (*9–11*). Although some models have been developed to guide the design of Boost converters (*12–16*), which could roughly predict dynamic behaviors including voltage overshoots, a critical gap remained between theoretical predictions and practical behavior. A primary deficiency of these models lies in their neglect of parasitic elements, such as the equivalent series resistance (ESR) of inductors and capacitors, and the non-ideal characteristics of semiconductor switches. It shows profound implications for the accuracy of the modeling and the reliability of the Boost converters (*8*). Furthermore, the prediction inaccuracies of conventional models became even more pronounced in advanced converter topologies, including interleaved and multiphase Boost configurations, or systems employing maximum power point tracking (MPPT) algorithms. In these settings, dynamic interactions among multiple switching phases, control loops, and load conditions introduced additional nonlinearity and coupling issues, further complicating accurate predictions (*17–20*).

To meet the growing demand for controlling voltage overshoots with high precision and optimizing overall system efficiency, accurate dynamic models are paramount. Without a detailed understanding of the dynamic interactions in Boost converters, engineers have to rely on empirical tuning and iterative prototyping methods that are resource-intensive and inefficient. Furthermore, such approaches may lead to significant overdesigns, resulting in higher costs, greater weights, and slower response times, which undermine the economic and physical performance of electrical systems (*21*). The combined effects of safety concerns and the drawbacks of overdesign (particularly concerning size and weight) substantially limit the applicability of Boost converters in high-reliability scenarios (such as aerospace electrical systems). As Boost converters become more integral to applications with stringent transient control requirements, it is critical to develop comprehensive analytical models with high accuracy to predict both steady-state and dynamic behaviors. First-principle modeling, widely applied in fields such as quantum mechanics, condensed-matter physics, and computational chemistry, has proven powerful in resolving fundamental problems by deriving system behavior directly from governing physical laws rather than empirical assumptions. Such approaches eliminate reliance on curve fitting or parameter heuristics, thereby ensuring both universality and predictive accuracy. However, despite its transformative impact in other scientific domains, first-principle modeling has rarely been applied to power electronics. Its introduction will offer a rigorous and universal framework to uncover system dynamics, enhance predictive accuracy, and enable the design of converters that meet the stringent demands of future high-reliability applications.

Herein, this study presented a first-principle modeling framework of Boost converters for the first time by deriving precise dynamic equations. We came up with the steady-state approximations and constructed two time-domain models that incorporated non-ideal component coupling. Compared to conventional Boost converter models, the steady-state errors between experimental output voltages and simulated voltages by the proposed models were reduced by a factor of 11.0 (from 20.9% to 1.9%, under input voltage variations) and by a factor of 10.2 (from 15.3% to 1.5%, under load variations), and the dynamic-state errors of output voltage overshoots were reduced by a factor of 15.4 (from 77.1% to 5.0%, under input voltage changes) and by a factor of 35.1 (from 42.1% to 1.2%, under load changes). Furthermore, we investigated the mechanisms between individual parameters and output voltages and delineated overshoot mitigation pathways for several representative scenarios. Building on this theoretical framework, we proposed practical, low-complexity methods to suppress voltage overshoot in Boost converters and validated them on a deployed spaceborne electrical system.

## First-principle modeling framework

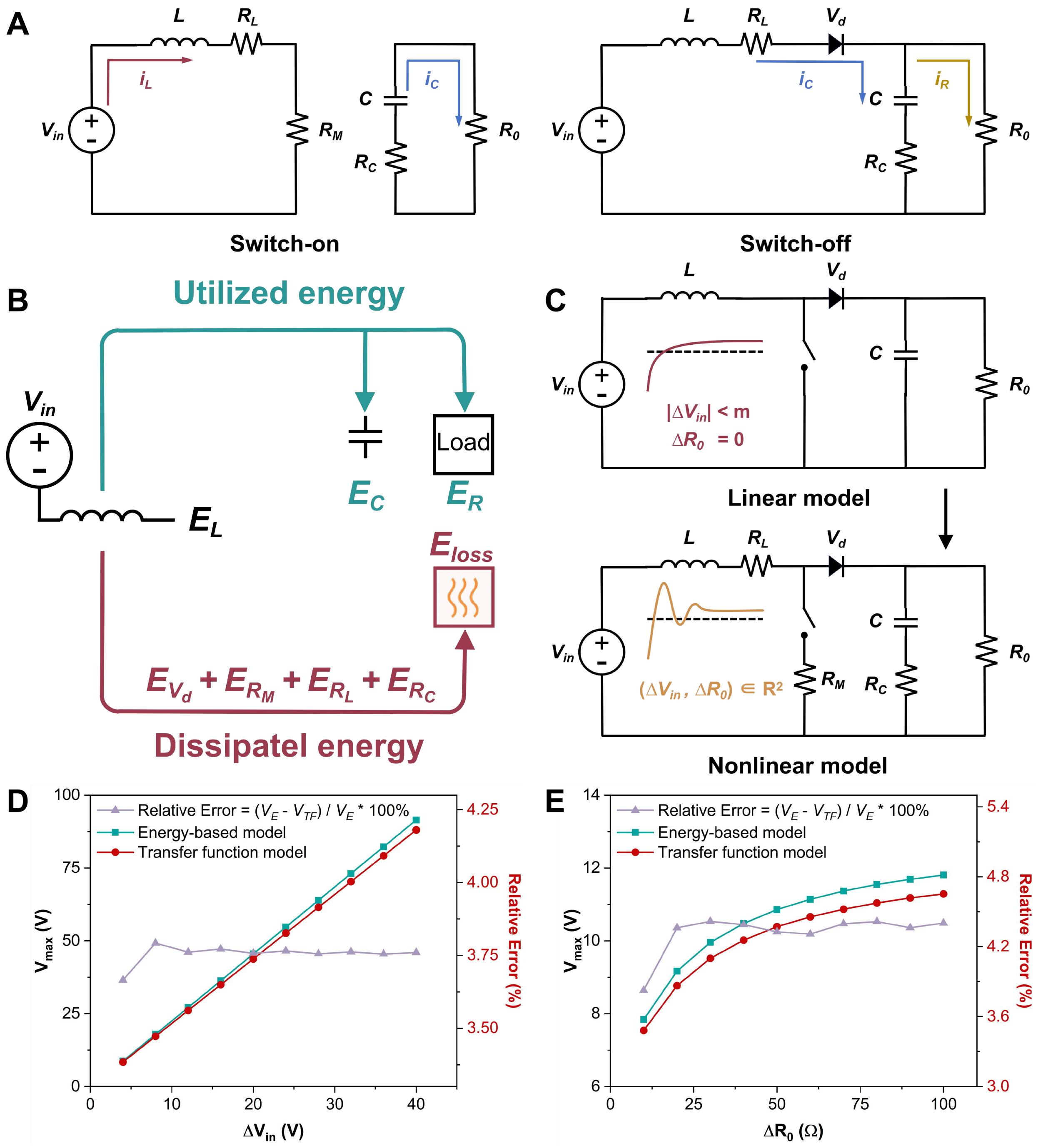

**Fig. 2. First-principle modeling framework of Boost converter dynamics.** (**A**) The Boost converter operates in distinct modes during switch-on and switch-off states. (**B**) In the proposed energy-based model, part of the energy released by the inductor is utilized by the capacitor and the load, while the remainder is dissipated through parasitic elements. Environmental disturbances disrupt this energy balance, leading to rapid fluctuations in the output voltage as the dependent variable. (**C**) Conventional linear transfer functions are valid only for small input voltage variations and constant load resistance, and thus fail to capture transient dynamics. The proposed nonlinear transfer function model remains applicable under arbitrary variations in input voltage and load resistance, enabling accurate prediction of transient behavior. (**D**) and (**E**) When varying the input voltage and load resistance, the relative error between the proposed two models in predicting output voltage overshoot remains below 5%.

When the MOSFET operates in different states (on or off), the Boost converter functions in distinct operating modes (**Fig. 2A**). When the switch is on, the power supply charges the inductor, while the capacitor discharges to provide energy to the load. When the switch is off, the inductor releases energy to charge the capacitor and power the load. Herein, we proposed two analytical models for the output voltage based on first-principles, considering both the law of energy conservation and the transfer function, and compared them for validation. Based on the law of energy conservation, the energy released by the inductor at any given time is utilized by the capacitor and the load, as well as dissipated through parasitic parameters (**Fig. 2B**), satisfying the following energy balance equation:

$$E_L = E_C + E_R + E_{loss,V_d} + E_{loss,R_M} + E_{loss,R_L} + E_{loss,R_C} \quad (1)$$

By combining equation (1) with the voltage relations among the circuit parameters, a second-order differential equation (2) for the output voltage is obtained, which can be transformed into a time-domain expression incorporating nonlinear transient behavior. The complete mathematical derivations are shown in Supplementary Equations (1)-(17).

$$LC \cdot \frac{d^2 v}{dt^2} + [\frac{L}{R_0} + C(R_L + DR_M)] \cdot \frac{dv}{dt} + [(1-D)^2 + \frac{(1-D)^2 R_C + R_L + DR_M}{R_0}] \cdot v = (1-D)V_{in} - (1-D)^2 V_d \quad (2)$$

According to energy balance equation (1), dynamic environments can disrupt energy conservation, causing a mismatch between energy injection and dissipation, which leads to rapid fluctuations in the output voltage as the dependent variable, resulting in voltage overshoot. The input voltage ($V_i$) controls inductor energy storage; a rapid increase can cause voltage overshoots if not compensated. Load resistance ($R_0$) affects capacitor discharge; a rapid increase in $R_0$ slows energy dissipation, raising output voltage. The duty cycle ($D$) defines inductor charging time; abrupt changes can mismatch energy delivery, causing overshoots. While $D$ can be gradually regulated, both $V_i$ and $R_0$ often change abruptly, for example during start-up transients or shifts in load operating modes, due to environmental and operational factors, making them primary sources of overshoot. From the energy storage perspective, the inductance ($L$) sets the rate of current change, where a larger $L$ delays current adaptation to disturbances, thereby amplifying overshoot. The output capacitor ($C$) smooths voltage, but insufficient capacitance fails to absorb sudden inductor discharges, while excess capacitance prolongs recovery. The parasitic resistance of inductor ($R_L$) and capacitor ($R_C$) introduces distributed energy dissipation, acting as intrinsic damping elements. Lower $R_L$ and $R_C$ reduce energy loss, allowing sharper current peaks that exacerbate voltage spikes. In addition, diode forward voltage drop ($V_d$) introduces a voltage threshold that delays inductor discharge into the output, leading to transient energy accumulation. A smaller $V_d$ increases the likelihood of overshoot in the output voltage at the onset of conduction. Lastly, MOSFET on-resistance ($R_M$) impacts charging efficiency, and lower $R_M$ reduces losses but accelerates current rise, increasing overshoot risk. These parameters interact nonlinearly under real-world disturbances, making idealized models inadequate. The proposed energy-based model incorporating parasitic parameters effectively addresses this challenge, offering strong physical interpretability and broad applicability.

In parallel, based on the voltage-current relations at the Boost converter nodes, we proposed a nonlinear transfer function model that accounts for abrupt variations in $V_i$ and $R_0$ as well as the influence of parasitic parameters (**Fig. 2C**). The transfer functions of the output voltage $V_o$ with respect to $V_i$ and $R_0$ are as follows, respectively:

$$G_s(\frac{V_o}{V_i}) = \frac{(1-D)CR_0R_C \cdot s + (1-D)R_0 - \frac{V_d}{V_i}(1-D)^2 R_0}{(R_0+R_C)LC \cdot s^2 + [(1-D)^2 CR_0R_C + DCR_M(R_0+R_C) + CR_L(R_0+R_C) + L] \cdot s + (1-D)^2 R_0 + R_L + DR_M + (1-D)^2 R_C} \quad (3)$$

$$G_s(\frac{V_o}{R_0}) = \frac{gs^4 + hs^3 + js^2 + ks + l}{as^4 + bs^3 + cs^2 + ds + f} \quad (4)$$

The values of all coefficients and the complete mathematical derivations can be found in Supplementary Equations (18)-(78) and **fig. S1**. Conventional linear transfer functions (**Fig. 2C**) neglect parasitic effects and approximate input voltage variations through small-signal models, thereby predicting the output voltage as an idealized, gradually rising curve. Such models also fail to capture the impact of load changes on output behavior. The proposed nonlinear transfer function model overcomes these limitations, remaining valid under arbitrary variations of input voltage and load resistance, and accurately predicts transient overshoots in the output voltage. To assess the reliability and consistency of the proposed two models and to strengthen their physical interpretation and theoretical completeness, we compared their predictions of voltage overshoot under variations in input voltage (**Fig. 2D**) and load resistance (**Fig. 2E**). Across all scenarios, as the disturbance intensity increased, the relative error between the two models remained below 5%, demonstrating both high reliability and strong consistency.

## Accurate prediction of output voltages

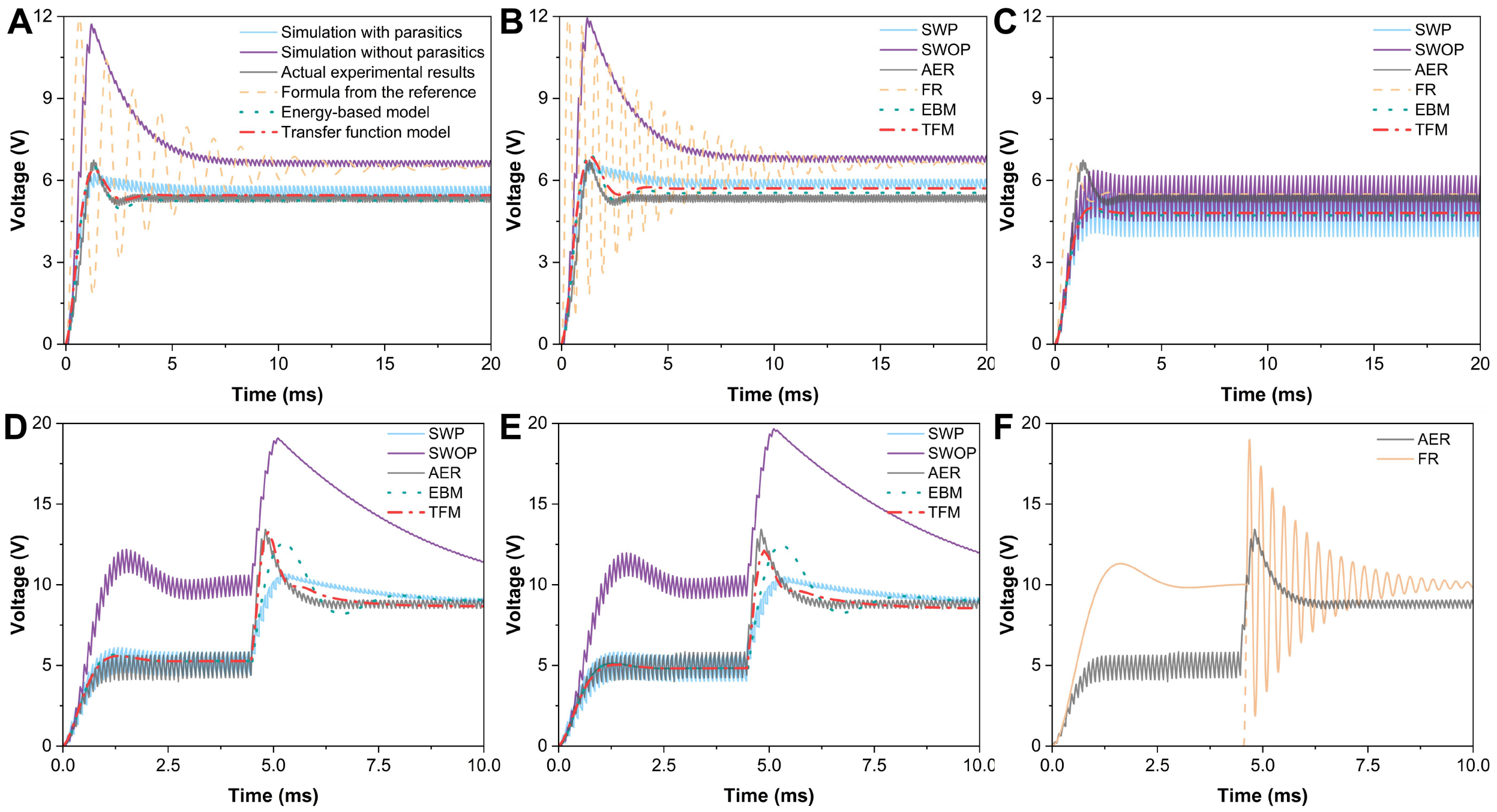

**Fig. 3. Comparisons of fitting results for different models.** (**A** to **C**) Accuracy comparisons for different methods under input voltage variations. (**A**) Curve fittings for different models are conducted by selecting appropriate values within the error range of the nominal component values. (**B**) Curve fittings for different models are performed using experimentally measured values. (**C**) Adjusting the FR to fit the measured waveform results in severe deviation from the actual physical configuration. (**D** to **F**) Accuracy comparisons for different methods under load resistance variations. (**D**) Curve fittings for different models are conducted by selecting appropriate values within the error range of the nominal value of the components. (**E**) Curve fittings for different models are performed using experimentally measured values. (**F**) The FR fails to capture the transient response under varying load conditions.

To evaluate the prediction accuracy of the proposed models, we conducted experiments involving changes in input voltage and load resistance separately and recorded the complete output voltage waveforms. The experimental and simulation groups included: simulation with parasitics (SWP), simulation without parasitics (SWOP), actual experimental results (AER), formula from the reference (FR) (*11*), energy-based model (EBM), and transfer function model (TFM). For the FR group, the transfer function of the Boost converter was derived by analogy to the transfer function of the Buck converter (*11*), followed by a transformation into the time domain for analysis (Supplementary Equations (79)-(94)). In comparison to the SWOP group, the SWP group incorporated parasitic resistances in series with the inductor and capacitor, and also accounted for the MOSFET on-resistance and the diode forward voltage drop. This study compared the fitting accuracies of three methods in predicting the output voltages during input voltage variations with system startup as an example (**Fig. 3, A** to **C**). The first method used component values within the error range of their nominal values, resulting in EBM and TFM curves closely matching the AER curve. The second method employed a high-precision LCR digital bridge to measure the components and plotted curves based on these measured values for comparison. **Table S1** provides the detailed parameters used in these methods. The results revealed that the accuracies of the proposed models were substantially superior to that of conventional models. The proposed models accurately fitted both the steady-state and dynamic responses of the output voltage waveforms. In the TFM group, the steady-state error of output voltage was reduced from 20.9% to 1.9% and the dynamic-state error of output voltage overshoot decreased from 77.1% to 5.0% (**Fig. 3A** and **table S2**). The third method used the FR curve as a reference, adjusting its component parameters to align closely with the AER curve. However, the inductance had to be set to 20 times its nominal value, and the capacitance to 1/8 of its nominal value, indicating a severe deviation from the actual physical configuration. These findings highlighted the numerical superiority and the strong physical interpretability and parameter consistency of the proposed models in this study. Moreover, the close alignment between the EBM, TFM, and ER curves across different component conditions demonstrated the robustness under practical tolerances.

Few previous studies have mentioned analytical equations for load resistance variations. This study compared the proposed models with actual physical measurements and simulations of the Boost converter to assess fitting accuracy (**Fig. 3, D** and **E**). The first method again used component values within their nominal error range, resulting in EBM and TFM curves closely matching the AER curve. The second method employed the high-precision LCR digital bridge for component parameter measurements and plotted corresponding curves for comparison. **Table S3** lists the detailed parameters for these methods. **Fig. 3F** illustrates the fitting accuracy of the FR under load variations, it failed to capture the transient responses. As a result, the waveform was reconstructed by stitching together fitted curves corresponding to two different load resistances. This method also employed the high-precision LCR digital bridge for component measurement. The results showed that the accuracies of the proposed models in this study were significantly superior to that of conventional models. Finally, in the TFM group, the steady-state error of output voltage was reduced from 15.3% to 1.5%, and the dynamic-state error of output voltage overshoot decreased from 42.1% to 1.2% (**Fig. 3D** and **table S4**). Unlike conventional models that typically rely on steady-state assumptions or parameter tuning post-simulation, the proposed models yield accurate predictions directly from circuit parameters without adjustments, even under rapidly changing load conditions. The observed reductions in both steady-state and transient errors underscore the potential of proposed models for real-time applications, where predictive accuracy under dynamic load conditions is critical.

## The influences of component parameters on output voltages

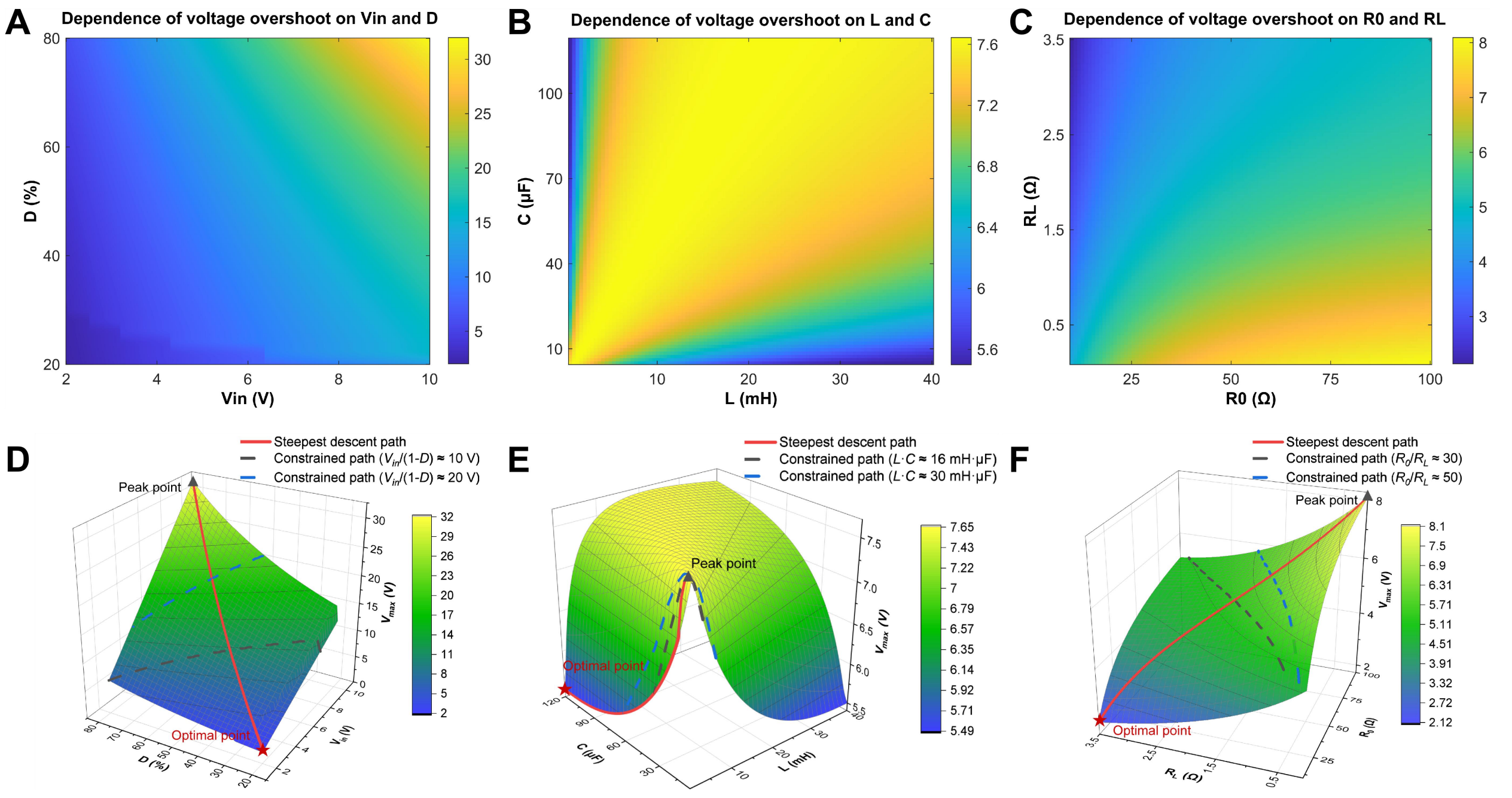

**Fig. 4. Influences of component parameters on voltage overshoots.** (**A**) As $V_i$ and $D$ increase, $V_{max}$ rises monotonically, with larger values concentrated in the region where both $V_i$ and $D$ are high. (**B**) Increases in $L$ and $C$ cause $V_{max}$ to rise initially and then decline, with the critical values of $L$ and $C$ determined by the system itself. (**C**) As $R_0$ increases and $R_L$ decreases, $V_{max}$ rises monotonically, with larger values concentrated in the region of high $R_0$ and low $R_L$. (**D** to **F**) Three-dimensional response surface of the peak voltage. The solid red line denotes the steepest descent path, and dashed lines denote optimization paths under different constraints. (**D**) When optimizing voltage overshoot by varying $V_i$ and $D$, the practical constraint is to maintain constant steady-state output voltage. (**E**) When varying $L$ and $C$, the constraint is to preserve the circuit's characteristic frequency. (**F**) When varying $R_0$ and $R_L$, the constraint is to keep parasitic losses small.

To achieve precise regulation of output voltages, it is crucial to investigate the effects of various component parameters on their behaviors. Herein, the start-up phase (a step change in $V_i$) was chosen as a representative case. By independently varying each parameter, we obtained the output voltage response mechanisms from physical experiments and different models (**fig. S2**), with the complete waveform for each point provided in **fig. S3 to S11**. **Table S5** illustrates the fitting errors of different models, with detailed histograms of the RMSE provided in **fig. S12** and **S13**. Based on the comparisons of fitting errors, it can be intuitively corroborated that the proposed models (taking the TFM model as an example) in this study exhibit a remarkable improvement in accuracy.

After validating the close agreement between the fitting results of the proposed TFM model and experimental measurements, continuous scanning was used to obtain heat maps of voltage overshoot under several representative parameter combinations (**Fig. 4, A** to **C**). As $V_i$ and $D$ increase, the peak voltage ($V_{max}$) rises monotonically, with larger values concentrated in the region where both $V_i$ and $D$ are high. Increases in $L$ and $C$ cause $V_{max}$ to rise initially and then decline, with the critical values of $L$ and $C$ determined by the system itself. As $R_0$ increases and $R_L$ decreases, $V_{max}$ rises monotonically, with larger values concentrated in the region of high $R_0$ and low $R_L$. These results indicated that input voltages, duty cycles, load resistances, and parasitic parameters jointly affect both the steady-state and dynamic behavior of the output voltages, though the response patterns remain simple. In contrast, passive storage elements (inductance and capacitance) introduced more complex and non-monotonic dynamics due to energy accumulation and response damping, while having little impact on the steady-state output. Building on the heat maps, we reconstructed the response surface in three-dimensional parameter space, enabling a more direct visualization of the descent pathways of the peak voltage (**Fig. 4, D** to **F**). By computing the gradient and local descent rate at each coordinate, we obtained the steepest descent paths (solid red lines). In practice, operational constraints often preclude following this path directly, for example, maintaining constant steady-state output voltage (**Fig. 4D**), preserving the circuit's characteristic frequency (**Fig. 4E**), or keeping parasitic losses small (**Fig. 4F**). Even under these constraints, the resulting constrained optimization paths (dashed lines) effectively reduce voltage overshoot, offering substantial flexibility and practical value.

**Strategies for mitigating voltage overshoots in space electrical systems**

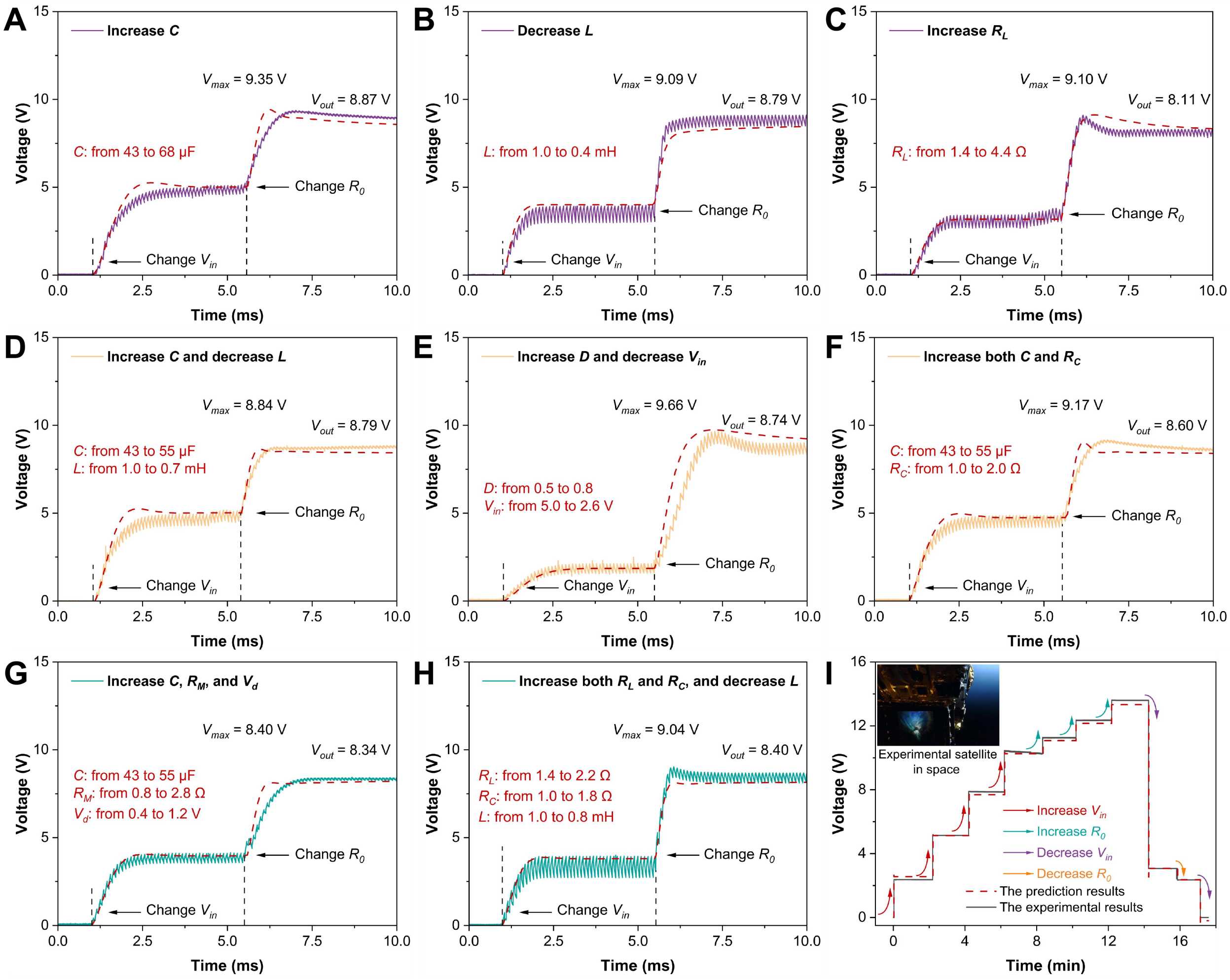

**Fig. 5. Strategies for mitigating or eliminating voltage overshoots.** (**A**) Altering $C$, (**B**) $L$, and (**C**) $R_L$ separately to mitigate output voltage overshoots. (**D**), (**E**), and (**F**) Simultaneously introducing minor adjustments to two components serves as a practical alternative to large-scale modification of a single component. (**G**) and (**H**) Simultaneous tuning of three parameters enables more flexible component selection and requires only minimal numerical adjustments. (**I**) Verification of the Boost system on the satellite based on the proposed models. By continuously varying input voltage and load resistance in the space environment, the output performances of the Boost converter were measured.

To mitigate or eliminate voltage overshoots (**Fig. 3D**), a range of parameters can be adjusted to achieve a controllable Boost effect (output voltage) in practical applications. As outlined in the preceding sections, this study analyzed the output voltage responses under varied component parameters. These analyses unveiled clear correlations between specific parameter modifications and distinct output voltage waveform behaviors, thereby establishing a strong theoretical foundation for systematic strategies in output voltage shaping. Before making physical adjustments, waveform predictions of TFM (represented by dashed lines in **Fig. 5**) were conducted to evaluate the suitability of the proposed parameter modifications by varying the input voltage from 0 V to 5 V and the load resistance from 10 Ω to 150 Ω independently. Except for a few instances where peak times exhibited minor discrepancies, the voltage overshoots and steady-state outputs demonstrated a high degree of accuracy in the predictions. The specific errors are detailed in **table S6**. These results substantiate the practical applicability of the proposed analytical models, allowing circuit behavior to be fully evaluated and optimized in the simulation stage before implementation. Building on these foundations, the controllable adjustment of output voltage was successfully achieved by modifying various parameter combinations, effectively mitigating voltage overshoots. Altering $C$ and $L$ separately can effectively mitigate dynamic voltage overshoots without affecting the steady-state outputs (**Fig. 5, A** and **B**), and adjusting the parasitic resistance $R_L$ individually reduces voltage overshoot at the cost of a slight degradation in steady-state performance (**Fig. 5C**). This finding confirmed that even single-parameter tuning can be sufficient to suppress transient anomalies under certain conditions. However, in real-world applications, components and their parasitic parameters often change simultaneously, and individual parameters may not allow for a wide range of adjustments. Consequently, voltage overshoots can be mitigated by tuning two or three parameters concurrently (**Fig. 5, D** to **H**). The combined effects of small changes in two parameters lead to effective mitigation of voltage overshoots and maintaining steady-state performances (**Fig. 5, D** and **E**). These results demonstrated that multi-parameter coordination can achieve a better balance between dynamic suppression and steady-state integrity than single parameter approaches. **Fig. 5F** demonstrates the effect of simultaneous adjustments of $C$ and $R_C$, which results in a much smaller loss of steady-state performance as compared to changes in $R_C$ alone. This result indicated that appropriate parameter pairing based on the proposed models can compensate for the limitations of individual components, improving both efficiency and stability. Further, adjusting three parameters simultaneously offers greater flexibility in component selection, with minimal variation in each component parameter (**Fig. 5, G** and **H**). Such multi-dimensional tuning not only expanded the design flexibility but also met system-level constraints (such as thermal, cost, and size/weight limits) without sacrificing converting performance. Consequently, the proposed analytical models serve as a powerful tool for precise and safe Boost converter designs.

The experimental results provide valuable theoretical insights for the design of efficient and reliable Boost converters, enabling flexible adjustment of component combinations to mitigate voltage overshoots and optimize the Boost effect even in extreme conditions. Furthermore, by analogy, the analytical models for various DC−DC converters can also be derived, facilitating high-precision control of output voltage characteristics. Based on the proposed analytical models for the Boost converter, this study designed a spaceborne Boost system, which has completed on-orbit deployment and function verification with a satellite (*24*, *25*). The input voltage was varied by adjusting the number of connected PV panels (approximately 2 V, 4 V, 6 V, and 8 V), and the load resistance was changed by controlling the operational mode of the load (approximately 25 Ω, 35 Ω, 65 Ω, and 300 Ω). Data transmitted from the satellite indicated that the Boost module designed according to the proposed models delivered a stable output voltage without significant voltage overshoots, and demonstrated high reliability and stability under dynamic conditions (**Fig. 5I** and **fig. S14**). This successful space deployment underscores the real-world applicability and robustness of the proposed models, bridging the gap between theoretical modeling and critical engineering environments such as aerospace power systems.

## Conclusions

This study introduces a first-principle modeling framework for Boost converters that captures dynamic behaviors under rapid input voltage and load variations by incorporating non-ideal component coupling. The proposed models significantly improves prediction accuracy, reduces voltage overshoots, and enhances system reliability in mission-critical applications (*26*–*31*). It bridges the gap between theoretical modelling and practical engineering constraints, enabling more efficient and robust converter designs. Extensible to other DC−DC converter topologies, this framework provides a foundation for the precise design of electrical converting systems across microelectronics, renewable energy, and aerospace applications. Future work could address additional practical factors, such as electromagnetic interference and parasitic coupling, particularly in high-frequency environments. Integration with real-time control algorithms and artificial intelligence may further enable adaptive optimisation under dynamic operating conditions, paving the way for the next generation of energy conversion technologies.

**Acknowledgments:**

We acknowledge financial support from the Fundamental Research Funds for the Central Universities (D5000220072) and the Northwestern Polytechnical University (23GH0202211, 24SH0201254). We acknowledge Shenzhen JLC Electronics Co., Ltd. for providing PCB manufacturing services. We acknowledge the Shaanxi Institute of Electronic Information Product Supervision and Inspection for providing reliability testing of the circuit system before space testing.

**Funding**

Fundamental Research Funds for the Central Universities D5000220072

Northwestern Polytechnical University 23GH0202211

Northwestern Polytechnical University 24SH0201254

**Author contributions**

T.Y. conceived the project and designed the research. T.Y., Y.W., W.L., Z.W., and X.Z. developed the theoretical framework and derived the analytical equations. J.S., Q.X., Z.G., and J.R. performed the simulations and experimental validations. Y.W., W.L., and T.Y. wrote the manuscript with input from all authors. T.Y. supervised the entire project. All authors discussed the results and contributed to the final manuscript.

**Competing interests**

The authors declare no competing interests.

**Data and materials availability**

Additional data that support the findings of this study are available from the corresponding author upon reasonable request.

**List of Supplementary Materials:**

Materials and Methods

Figs. S1 to S14

Tables S1 to S6

# Supplementary Materials

## First-Principle Modeling Framework of Boost Converter Dynamics for Precise Energy Conversions in Space

Yifan Wang[1], Wenhua Li[1], Zhenlong Wang[1], Xinrui Zhang[1], Jianfeng Sun[1], Qianfu Xia[1], Zhongtao Gou[1], Jiangang Rong[2], Tao Ye[1]*

[1]Ministry of Education Key Laboratory of Micro/Nano Systems for Aerospace, Key Laboratory of Micro- and Nano-Electro-Mechanical Systems of Shaanxi Province, School of Mechanical Engineering, Northwestern Polytechnical University; Xi'an, 710000, China.

[2]Shanghai Guoyu Zhilian Aerospace Technology Co., Ltd; Shanghai, 201112, China.

*Corresponding author. Email: yetao@nwpu.edu.cn

## Materials and Methods

### Energy-based model

### A. Energy-landscape perturbation–response mechanism in Boost transients

When an external perturbation disrupts the steady-state energy balance of a boost converter, the output voltage responds with a rapid excursion. During the interval from the onset of the disturbance to the attainment of the peak voltage, the circuit obeys the energy-conservation relation:

$$E_L = E_C + E_R + E_{loss,V_d} + E_{loss,R_M} + E_{loss,R_L} + E_{loss,R_C} \tag{1}$$

$$\begin{aligned}\frac{1}{2}L(i_{L,peak}^2 - i_{L,t_0}^2) = \frac{1}{2}C(V_{\max}^2 - V_{t_0}^2) + \int_{t_0}^{t_1}\frac{V_{out}^2(t)}{R_0}\cdot dt + \int_{t_0}^{t_1} i_D(t)V_d \cdot dt + D\int_{t_0}^{t_1} i_L^2(t)R_M \cdot dt \\ + \int_{t_0}^{t_1} i_L^2(t)R_L \cdot dt + \int_{t_0}^{t_1} i_C^2(t)R_C \cdot dt\end{aligned} \tag{2}$$

The left-hand side of equation (1) represents the magnetic energy released by the inductor, whereas the first term on the right-hand side corresponds to the electrostatic energy absorbed by the output capacitor. The second term quantifies the energy dissipated in the load, and the remaining terms account for losses associated with parasitic parameters.

As the input voltage ($V_i$) and the duty cycle ($D$) increase, equation (3) becomes larger, causing the peak voltage ($V_{max}$) to rise.

$$i_{L,peak} = i_{L,t_0} + D(t_1 - t_0)\frac{V_i - i_L R_L - i_L R_M}{L} \tag{3}$$

As the load resistance ($R_0$) increases, $E_R$ decreases, causing $V_{max}$ to rise. When the inductance ($L$) is small, the rapid current decay during the off-phase results in incomplete energy transfer to the capacitor, with a substantial portion of the stored magnetic energy dissipated as heat. Increasing $L$ moderates the discharge rate, enhances energy delivery to the capacitor, and thus raises $V_{max}$. However, as $L$ becomes large, further increases reduce the quantity defined by equation (4), thereby leading to a decline in $V_{max}$.

$$\begin{aligned} E_L &= \frac{1}{2}L(i_{L,peak}^2 - i_{L,t_0}^2) \\ &= \frac{1}{2}\frac{D^2(V_i - i_L R_L - i_L R_M)^2(t_1 - t_0)^2}{L} + i_{L,t_0}D(V_i - i_L R_L - i_L R_M)(t_1 - t_0) \end{aligned} \tag{4}$$

Similarly, when the capacitance ($C$) is small, its limited charge-storage capacity leads to incomplete absorption of the delivered energy, with significant losses manifesting as heat. As $C$ increases, energy absorption becomes more efficient, resulting in a higher $V_{max}$. However, once $C$ exceeds a critical threshold, further increases reduce $V_{max}$ as governed by the $E_C$ term in equation (2). Moreover, as $V_d$, $R_M$, $R_L$, and $R_C$ increase, the corresponding loss terms in equation (2) grow accordingly, thereby diminishing the net energy delivered to the output stage and resulting in a reduction of $V_{max}$.

**B. Equation based on the law of energy conservation**

During the transition from disturbance to the reestablishment of steady state, the energy conservation equation can be written as:

$$E_L + E_C = E_{in} - E_R - E_{loss} \tag{5}$$

Differentiation of equation (5) with respect to time yields the instantaneous power balance:

$$Li_L \frac{di_L}{dt} + Cv\frac{dv}{dt} = V_{in} i_L - \frac{v^2}{R_0} - (R_L + DR_M){i_L}^2 - (1-D)V_d i_L - {i_C}^2 R_C \tag{6}$$

According to Kirchhoff's Voltage Law (KVL), the voltage relationship can be expressed as:

$$L\frac{di_L}{dt} = V_{in} - (1-D)(v + V_d) - i_L(R_L + DR_M) \tag{7}$$

By substituting equation (7) into equation (6), we obtain:

$$Cv\frac{dv}{dt} = (1-D)vi_L - \frac{v^2}{R_0} - {i_C}^2 R_C \tag{8}$$

By neglecting the second-order term ${i_C}^2 R_C$, the solution is obtained as:

$$i_L = \frac{C\frac{dv}{dt} + \frac{v}{R_0}}{1-D} \tag{9}$$

$$\frac{di_L}{dt} = \frac{C\frac{d^2 v}{dt^2} + \frac{\frac{dv}{dt}}{R_0}}{1-D} \tag{10}$$

By substituting equations (9) and (10) into equation (7) and simplifying, the second-order differential equation is obtained as:

$$LC \cdot \frac{d^2 v}{dt^2} + [\frac{L}{R_0} + C(R_L + DR_M)] \cdot \frac{dv}{dt} + [(1-D)^2 + \frac{R_L + DR_M}{R_0}] \cdot v = (1-D)V_{in} - (1-D)^2 V_d \tag{11}$$

Due to the non-transient, sustained loss associated with $R_C$, which primarily affects the steady-state behavior of the output voltage, and considering that both $R_C$ and $R_0$ contribute as resistive elements at the output port, the following correction is introduced to account for the neglected $R_C$ loss:

$$(1-D)^2+\frac{R_L+DR_M}{R_0}\Rightarrow\frac{(1-D)^2R_0+(1-D)^2R_C+R_L+DR_M}{R_0} \tag{12}$$

The complete second-order differential equation is expressed as:

$$LC\cdot\frac{d^2v}{dt^2}+[\frac{L}{R_0}+C(R_L+DR_M)]\cdot\frac{dv}{dt}+[(1-D)^2+\frac{(1-D)^2R_C+R_L+DR_M}{R_0}]\cdot v=(1-D)V_{in}-(1-D)^2V_d \tag{13}$$

By comparing with the standard form of equation (14), the time-domain expression for the output voltage is obtained as shown in equation (15):

$$\frac{d^2v}{dt^2}+2\xi\omega_0\cdot\frac{dv}{dt}+\omega_0{}^2\cdot v=\omega_0{}^2V_\infty \tag{14}$$

$$\begin{aligned}v(t)=V_\infty+[v(0)-V_\infty]\cdot e^{-\xi\omega_0 t}\cdot[\cos(\omega_d t)+\frac{\xi}{\sqrt{1-\xi^2}}\sin(\omega_d t)]\\+\frac{e^{-\xi\omega_0 t}}{\omega_d}\cdot[\frac{dv(0)}{dt}+\xi\omega_0(v(0)-V_\infty)]\cdot\sin(\omega_d t)\end{aligned} \tag{15}$$

$$\omega_d=\omega_0\sqrt{1-\xi^2} \tag{16}$$

$$\frac{dv(0)}{dt}=\frac{2v(0)}{C}\cdot(\frac{1}{R_1}-\frac{1}{R_2}) \tag{17}$$

Under zero conditions, the last term of equation (15) vanishes. In equation (17), $R_1$ and $R_2$ represent the load resistances before and after the change, respectively.

## Transfer function model

### A. Steady-state analysis under varied input voltages

In a single cycle $T$, when the MOS transistor is either on or off, the circuit operates in different states. Under steady-state conditions, the inductor balances energy storage and discharge. Therefore, the volt-second balance principle is applied, expressed as:

$$(V_i - I_{L1}R_L - I_{L1}R_M)DT = (V_o + I_{L2}R_L + V_d - V_i)(1-D)T \tag{18}$$

where $V_i$ is the input voltage, $V_o$ is the output voltage, and $D$ is the duty cycle of the PWM signal, $I_{L1} = I_{L2} = I_L$ are the average currents of the inductor during the charging and discharging processes, respectively. This equation simplifies to:

$$V_i = (1-D)V_o + I_L R_L + DI_L R_M + (1-D)V_d \tag{19}$$

Similarly, the ampere-second balance principle is applied to the capacitor, given by:

$$\frac{V_o}{R_0}DT = (I_{L2} - \frac{V_o}{R_0})(1-D)T \tag{20}$$

where $R_0$ is the load resistance, and it simplifies to:

$$\frac{V_o}{R_0} = I_L(1-D) \tag{21}$$

By combining equations (19) and (21) and eliminating $I_L$, the steady-state analytical equation for the Boost converter can be derived:

$$V_o = \frac{[V_i - (1-D)V_d](1-D)R_0}{(1-D)^2 R_0 + R_L + DR_M} \tag{22}$$

$$I_L = \frac{V_i - (1-D)V_d}{(1-D)^2 R_0 + R_L + DR_M} \tag{23}$$

**B. Dynamic-state analysis under varied input voltages**

To analyze the dynamic behavior of the output voltage at system startup, it is essential to derive the transfer function of $V_o$ with respect to $V_i$. In this process, the average value of each parameter over a single switching period is considered. When $V_i$ is stable, the following relationships hold:

$$I_1(\mathrm{t}) = I_1 \tag{24}$$

$$I_2(\mathrm{t}) = I_2 \tag{25}$$

$$V_1(\mathrm{t}) = V_1 \tag{26}$$

$$V_2(\mathrm{t}) = V_2 \tag{27}$$

$$V_i(t) = V_i \tag{28}$$

When $V_i$ changes:

$$I_1(\mathrm{t}) = I_1 + \Delta I_1(\mathrm{t}) \tag{29}$$

$$I_2(\mathrm{t}) = I_2 + \Delta I_2(\mathrm{t}) \tag{30}$$

$$V_1(\mathrm{t}) = V_1 + \Delta V_1(\mathrm{t}) \tag{31}$$

$$V_2(\mathrm{t}) = V_2 + \Delta V_2(\mathrm{t}) \tag{32}$$

$$V_i(t) = V_i + \Delta V_i(t) \tag{33}$$

For $V_1(t)$, in time $DT$ and $(1 - D)T$:

$$V_1(\mathrm{t}) = I_1(\mathrm{t}) R_M \tag{34}$$

$$V_1(\mathrm{t}) = V_2(\mathrm{t}) + V_{\mathrm{d}} \tag{35}$$

Combining equations (34) and (35), we obtain:

$$V_1(\mathrm{t}) = (1 - D)(V_2(\mathrm{t}) + V_{\mathrm{d}}) + D I_1(\mathrm{t}) R_M \tag{36}$$

Similarly, for $I_2(t)$, in time $DT$ and $(1 - D)T$:

$$I_2(\mathrm{t}) = 0 \tag{37}$$

$$I_2(\mathrm{t}) = I_1(\mathrm{t}) \tag{38}$$

Combining equations (37) and (38), we derive:

$$I_2(\mathrm{t}) = (1 - D) I_1(\mathrm{t}) \tag{39}$$

When $V_i$ is stable:

$$V_1 = (1 - D)(V_2 + V_{\mathrm{d}}) + D I_1 R_M \tag{40}$$

$$I_2 = (1 - D) I_1 \tag{41}$$

When $V_i$ changes:

$$V_1 + \Delta V_1(\mathrm{t}) = (1-D)(V_2 + \Delta V_2(\mathrm{t}) + V_\mathrm{d}) + D(I_1 + \Delta I_1(\mathrm{t}))R_M \tag{42}$$

$$I_2 + \Delta I_2(\mathrm{t}) = (1-D)(I_1 + \Delta I_1(\mathrm{t})) \tag{43}$$

By subtracting equations (42) from (40) and (43) from (41), we obtain:

$$\Delta V_1(\mathrm{t}) = (1-D)\Delta V_2(\mathrm{t}) + D\Delta I_1(\mathrm{t})R_M \tag{44}$$

$$\Delta I_2(\mathrm{t}) = (1-D)\Delta I_1(\mathrm{t}) \tag{45}$$

Based on the voltage-current relationship, we derive the following:

$$\Delta V_i = \Delta V_1 + \Delta I_1 R_L + sL\Delta I_1 \tag{46}$$

$$\Delta V_o = \Delta V_2 = \Delta I_2 \frac{sCR_0R_C + R_0}{sC(R_0 + R_C) + 1} \tag{47}$$

Substituting equations (44) and (45) into equations (46) and (47) yields:

$$\begin{aligned} G(s) &= \frac{\Delta V_o}{\Delta V_i} \\ &= \frac{(1-D)CR_0R_C \cdot s + (1-D)R_0}{(R_0+R_C)LC \cdot s^2 + [(1-D)^2 CR_0R_C + DCR_M(R_0+R_C) + CR_L(R_0+R_C) + L] \cdot s + (1-D)^2 R_0 + R_L + DR_M} \end{aligned} \tag{48}$$

When the input voltage is modeled as a step signal, the steady-state voltage is given by:

$$\begin{aligned} C(\infty) &= \lim_{s\to 0} s\frac{V_i}{s}G(s) \\ &= V_i \frac{(1-D)R_0}{(1-D)^2 R_0 + R_L + DR_M} \end{aligned} \tag{49}$$

This is corrected in combination with the steady-state analytic equation derived earlier:

$$C(\infty) = V_i \frac{(1-D)R_0}{(1-D)^2 R_0 + R_L + DR_M} \Rightarrow C(\infty) = V_i \frac{(1-D)R_0 - \frac{V_d}{V_i}(1-D)^2 R_0}{(1-D)^2 R_0 + R_L + DR_M} \tag{50}$$

Further physical verification reveals that an increase in the parasitic resistance of the capacitor ($R_C$) leads to a decrease in the steady-state voltage. To account for this effect, an additional term, $mR_C$, is introduced in the denominator, with $m$ assumed to be $1 - D/D/(1 -$ $D)^2/D^2/1$. Through empirical fitting, it is found that when $m = (1-D)^2$, the fitting accuracy is maximized. Thus, the steady-state voltage is further refined as:

$$C(\infty) = V_i \frac{(1-D)R_0 - \frac{V_d}{V_i}(1-D)^2 R_0}{(1-D)^2 R_0 + R_L + DR_M} \Rightarrow C(\infty) = V_i \frac{(1-D)R_0 - \frac{V_d}{V_i}(1-D)^2 R_0}{(1-D)^2 R_0 + R_L + DR_M + (1-D)^2 R_C} \quad (51)$$

Therefore, the final transfer function is:

$$G(s) = \frac{(1-D)CR_0R_C \cdot s + (1-D)R_0 - \frac{V_d}{V_i}(1-D)^2 R_0}{(R_0+R_C)LC \cdot s^2 + [(1-D)^2 CR_0R_C + DCR_M(R_0+R_C) + CR_L(R_0+R_C) + L] \cdot s + (1-D)^2 R_0 + R_L + DR_M + (1-D)^2 R_C} \quad (52)$$

To simplify the calculation, $a$, $b$, $c$, $d$, and $f$ are used instead of the coefficients in (52). For step signal $\frac{k}{s}$, the output signal is:

$$\begin{aligned} C(s) &= R(s)G(s) \\ &= \frac{k}{s} \cdot \frac{ds+f}{as^2+bs+c} \\ &= k\left(\frac{f}{c} \cdot \frac{1}{s} + \frac{-dx_1+f}{ax_1^2-c} \cdot \frac{1}{s+x_1} + \frac{-dx_2+f}{ax_2^2-c} \cdot \frac{1}{s+x_2}\right) \end{aligned} \quad (53)$$

where $x_1 + x_2 = \frac{b}{a}$ and $x_1 x_2 = \frac{c}{a}$ are satisfied, and $a$, $b$, $c$, $d$, and $f$ are all greater than zero. Transforming the output signal, we obtain the time-domain analytical equation for the output voltage:

$$C(t) = k\left(\frac{f}{c} + \frac{-dx_1+f}{ax_1^2-c} \cdot e^{-x_1 t} + \frac{-dx_2+f}{ax_2^2-c} \cdot e^{-x_2 t}\right) \quad (54)$$

Through extensive calculations, it is determined that $x_1$ and $x_2$ form a pair of conjugate complex roots, while $\frac{-dx_1+f}{ax_1{}^2-c}$ and $\frac{-dx_2+f}{ax_2{}^2-c}$ form another pair of conjugate complex roots. The following parameters are introduced:

$$A = \frac{f}{2c} \quad (55)$$

$$B = \frac{bf - 2cd}{2c\sqrt{4ac - b^2}} \tag{56}$$

$$E = \frac{b}{2a} \tag{57}$$

$$F = \frac{\sqrt{4ac - b^2}}{2a} \tag{58}$$

Substituting these into equation (54) yields:

$$\begin{aligned} C(t) &= k[\frac{f}{c} + (-A + Bi) \cdot e^{(-E+Fi)t} + (-A - Bi) \cdot e^{(-E-Fi)t}] \\ &= k[\frac{f}{c} - 2\sqrt{A^2 + B^2} \cdot e^{-Et} \cdot \sin(Ft + \varphi)] \end{aligned} \tag{59}$$

where $\varphi = tan^{-1}(\frac{A}{B})$. By allowing time in equation (59) to approach infinity and setting its first derivative to zero, the steady-state and dynamic characteristics of the output voltage can be determined as follows:

$$V_o = V_i \frac{(1-D)R_0 - \frac{V_d}{V_i}(1-D)^2 R_0}{(1-D)^2 R_0 + R_L + DR_M + (1-D)^2 R_C} \tag{60}$$

$$t_p = \frac{\tan^{-1}(\frac{\sqrt{4ac-b^2}}{b}) - \tan^{-1}(\frac{f\sqrt{4ac-b^2}}{bf-2cd}) + \pi}{\frac{\sqrt{4ac-b^2}}{2a}} \tag{61}$$

$$V_{\max} = V_i[\frac{f}{c} - 2\sqrt{\frac{af^2 - bdf + cd^2}{4ac^2 - b^2c}} \cdot e^{\frac{b[\tan^{-1}(\frac{f\sqrt{4ac-b^2}}{bf-2cd}) - \tan^{-1}(\frac{\sqrt{4ac-b^2}}{b}) - \pi]}{\sqrt{4ac-b^2}}} \cdot \sin(\tan^{-1}(\frac{\sqrt{4ac-b^2}}{b}) + \pi)] \tag{62}$$

**C. Equation for varied load resistances**

When the load resistance changes, the output voltage consists of two components: the steady-state value of the previous moment $V_0$ and the amount of change $\Delta V_0$. When $R_0$ is stable:

$$R_0(t) = R_0 \tag{63}$$

When $R_0$ changes:

$$R_0(t) = R_0 + \Delta R_0(t) \tag{64}$$

By combining equations (63) and (64) with equations (24)-(27) and (29)-(32), we obtain:

$$\Delta V_1(\mathrm{t}) = (1-D)\Delta V_2(\mathrm{t}) + D\Delta I_1(\mathrm{t})R_M \tag{65}$$

$$\Delta I_2(\mathrm{t}) = (1-D)\Delta I_1(\mathrm{t}) \tag{66}$$

To simplify the calculation, $C$, $R_C$, and $R_0$ are considered as the overall impedance $Z$. Based on the voltage-current relationship, we obtain:

$$\Delta V_i = \Delta V_1 + \Delta I_1 R_L + sL\Delta I_1 = 0 \tag{67}$$

$$\Delta V_o = \Delta V_2 = I_2\Delta Z + \Delta I_2(Z + \Delta Z) \tag{68}$$

where $I_2$ can be calculated from the evolution of the result in equation (60):

$$\begin{aligned} I_2 &= (1-D)I_1 \\ &= \frac{(1-D)V_i - (1-D)^2 V_d}{(1-D)^2 R_0 + R_L + DR_M + (1-D)^2 R_C} \end{aligned} \tag{69}$$

Furthermore:

$$Z = \frac{sCR_0R_C + R_0}{sC(R_0 + R_C) + 1} \tag{70}$$

$$Z + \Delta Z = \frac{sC(R_0 + \Delta R_0)R_C + R_0 + \Delta R_0}{sC(R_0 + \Delta R_0 + R_C) + 1} \tag{71}$$

By combining equations (65), (66), (67), and (68), we derive:

$$\frac{\Delta V_o}{\Delta Z} = I_2 \frac{DR_M + R_L + sL}{DR_M + R_L + sL + (1-D)^2(Z + \Delta Z)} \tag{72}$$

By combining equations (70) and (71), we derive:

$$\frac{\Delta Z}{\Delta R_0} = \frac{(sCR_C + 1)^2}{[sC(R_0 + \Delta R_0 + R_C) + 1] \cdot [sC(R_0 + R_C) + 1]} \tag{73}$$

Therefore, the transfer function is:

$$
\begin{aligned}
G(s) &= \frac{\Delta V_o}{\Delta R_0} \\
&= \frac{\Delta V_o}{\Delta Z} \cdot \frac{\Delta Z}{\Delta R_0} \\
&= I_2 \frac{gs^4 + hs^3 + js^2 + ks + l}{as^4 + bs^3 + cs^2 + ds + f}
\end{aligned} \tag{74}
$$

The coefficients are as follows:

$$
\left\{
\begin{aligned}
& a = LC^3(R_0 + \Delta R_0 + R_C)^2(R_0 + R_C) \\
& b = LC^2(R_0 + \Delta R_0 + R_C)^2 + 2LC^2(R_0 + \Delta R_0 + R_C)(R_0 + R_C) \\
& \quad + C^3(DR_M + R_L)(R_0 + \Delta R_0 + R_C)^2(R_0 + R_C) \\
& \quad + C^3(1-D)^2(R_0 + \Delta R_0 + R_C)(R_0 + \Delta R_0)(R_0 + R_C)R_C \\
& c = 2LC(R_0 + \Delta R_0 + R_C) + LC(R_0 + R_C) \\
& \quad + C^2(DR_M + R_L)(R_0 + \Delta R_0 + R_C)(2R_0 + \Delta R_0 + 2R_C) \\
& \quad + C^2(DR_M + R_L)(R_0 + \Delta R_0 + R_C)(R_0 + R_C) \\
& \quad + C^2(1-D)^2(R_0 + \Delta R_0 + R_C)(R_0 + \Delta R_0)(R_0 + 2R_C) \\
& \quad + C^2(1-D)^2(R_0 + \Delta R_0)(R_0 + R_C)R_C \\
& d = L + C(DR_M + R_L)(3R_0 + 2\Delta R_0 + 3R_C) + C(1-D)^2(2R_0 + \Delta R_0 + 3R_C)(R_0 + \Delta R_0) \\
& f = (DR_M + R_L) + (1-D)^2(R_0 + \Delta R_0) \\
& g = LC^3(R_0 + \Delta R_0 + R_C){R_C}^2 \\
& h = LC^2{R_C}^2 + 2LC^2(R_0 + \Delta R_0 + R_C)R_C + C^3(DR_M + R_L)(R_0 + \Delta R_0 + R_C){R_C}^2 \\
& j = 2LCR_C + LC(R_0 + \Delta R_0 + R_C) + C^2(DR_M + R_L){R_C}^2 \\
& \quad + 2C^2(DR_M + R_L)(R_0 + \Delta R_0 + R_C)R_C \\
& k = L + C(DR_M + R_L)(R_0 + \Delta R_0 + 3R_C) \\
& l = DR_M + R_L
\end{aligned}
\right\} \tag{75}
$$

Due to computational errors in the derivation process, certain coefficients of the transfer function are adjusted based on empirical correction equations. The revised results are as follows:

$$\left\{\begin{aligned}
&a'=[0.5+(\frac{C-0.000042}{0.00005}+\frac{0.001-L}{0.0006}+\frac{R_L-1.4}{6}+\frac{D-0.5}{0.5}+\frac{R_C-1}{2}+\frac{V_d-0.4}{3}+\frac{R_M-0.8}{5})\times 0.6]\\
&\quad\times[LC^3(R_0+\Delta R_0+R_C)^2(R_0+R_C)]\\
&b'=[0.5+(\frac{C-0.000042}{0.00005}+\frac{0.001-L}{0.0006}+\frac{R_L-1.4}{6}+\frac{D-0.5}{0.5}+\frac{R_C-1}{2}+\frac{V_d-0.4}{3}+\frac{R_M-0.8}{5})\times 0.6]\\
&\quad\times[LC^2(R_0+\Delta R_0+R_C)^2+2LC^2(R_0+\Delta R_0+R_C)(R_0+R_C)\\
&\quad+C^3(DR_M+R_L)(R_0+\Delta R_0+R_C)^2(R_0+R_C)\\
&\quad+C^3(1-D)^2(R_0+\Delta R_0+R_C)(R_0+\Delta R_0)(R_0+R_C)R_C]\\
&c'=[0.5+(\frac{C-0.000042}{0.00005}+\frac{0.001-L}{0.0006}+\frac{R_L-1.4}{6}+\frac{D-0.5}{0.5}+\frac{R_C-1}{2}+\frac{V_d-0.4}{3}+\frac{R_M-0.8}{5})\times 0.6]\\
&\quad\times[2LC(R_0+\Delta R_0+R_C)+LC(R_0+R_C)+C^2(DR_M+R_L)(R_0+\Delta R_0+R_C)(2R_0+\Delta R_0+2R_C)\\
&\quad+C^2(DR_M+R_L)(R_0+\Delta R_0+R_C)(R_0+R_C)+C^2(1-D)^2(R_0+\Delta R_0+R_C)(R_0+\Delta R_0)(R_0+2R_C)\\
&\quad+C^2(1-D)^2(R_0+\Delta R_0)(R_0+R_C)R_C]\\
&d'=[0.5+(\frac{C-0.000042}{0.00005}+\frac{0.001-L}{0.0006}+\frac{R_L-1.4}{6}+\frac{D-0.5}{0.5}+\frac{R_C-1}{2}+\frac{V_d-0.4}{3}+\frac{R_M-0.8}{5})\times 0.6]\\
&\quad\times[L+C(DR_M+R_L)(3R_0+2\Delta R_0+3R_C)+C(1-D)^2(2R_0+\Delta R_0+3R_C)(R_0+\Delta R_0)]\\
&f=(DR_M+R_L)+(1-D)^2(R_0+\Delta R_0)\\
&g=LC^3(R_0+\Delta R_0+R_C){R_C}^2\\
&h=LC^2{R_C}^2+2LC^2(R_0+\Delta R_0+R_C)R_C+C^3(DR_M+R_L)(R_0+\Delta R_0+R_C){R_C}^2\\
&j=2LCR_C+LC(R_0+\Delta R_0+R_C)+C^2(DR_M+R_L){R_C}^2+2C^2(DR_M+R_L)(R_0+\Delta R_0+R_C)R_C\\
&k=L+C(DR_M+R_L)(R_0+\Delta R_0+3R_C)\\
&l=DR_M+R_L
\end{aligned}\right\} \tag{76}$$

Further:

$$\begin{aligned}C(s)&=R(s)G(s)\\&=\frac{\Delta R_0}{s}\cdot I_2\cdot\frac{gs^4+hs^3+js^2+ks+l}{as^4+bs^3+cs^2+ds+f}\end{aligned} \tag{77}$$

The final analytical equation for the output voltage is:

$$\begin{aligned}V_o'&=V_o+\Delta V_o\\&=V_i\cdot\frac{(1-D)R_0-\frac{V_d}{V_i}(1-D)^2R_0}{(1-D)^2R_0+R_L+DR_M+(1-D)^2R_C}+I_2\cdot\Delta R_0\cdot(E+F\cdot e^{3x_1\cdot t}+G\cdot e^{3x_2\cdot t}+H\cdot e^{3x_3\cdot t}+I\cdot e^{3x_4\cdot t})\end{aligned} \tag{78}$$

where $E$, $F$, $G$, $H$, $I$, $x_1$, $x_2$, $x_3$ and $x_4$ are the parameters derived from the time-domain conversion of equation (77).

**The formula of reference**

For $V_1(t)$, in time $DT$ and $(1-D)T$:

$$V_1(\mathrm{t}) = 0 \tag{79}$$

$$V_1(\mathrm{t}) = V_2(\mathrm{t}) \tag{80}$$

Combining equations (79) and (80), we obtain:

$$V_1(\mathrm{t}) = (1-D)V_2(\mathrm{t}) \tag{81}$$

Similarly, for $I_2(t)$, in time $DT$ and $(1-D)T$:

$$I_2(\mathrm{t}) = 0 \tag{82}$$

$$I_2(\mathrm{t}) = I_1(\mathrm{t}) \tag{83}$$

Combining equations (82) and (83), we derive:

$$I_2(\mathrm{t}) = (1-D)I_1(\mathrm{t}) \tag{84}$$

When $V_i$ is stable:

$$V_1 = (1-D)V_2 \tag{85}$$

$$I_2 = (1-D)I_1 \tag{86}$$

When $V_i$ changes:

$$V_1 + \Delta V_1(\mathrm{t}) = (1-D)(V_2 + \Delta V_2(\mathrm{t})) \tag{87}$$

$$I_2 + \Delta I_2(\mathrm{t}) = (1-D)(I_1 + \Delta I_1(\mathrm{t})) \tag{88}$$

By subtracting equations (87) from (85) and (88) from (86), we obtain:

$$\Delta V_1(\mathrm{t}) = (1-D)\Delta V_2(\mathrm{t}) \tag{89}$$

$$\Delta I_2(\mathrm{t}) = (1-D)\Delta I_1(\mathrm{t}) \tag{90}$$

Based on the voltage-current relationship, we derive the following:

$$\Delta V_i = \Delta V_1 + sL\Delta I_1 \tag{91}$$

$$\Delta V_o = \Delta V_2 = \Delta I_2 \frac{R_0}{sCR_0 + 1} \tag{92}$$

Substituting equations (89) and (90) into equations (91) and (92) yields:

$$G(s)=\frac{\Delta V_o}{\Delta V_i} = \frac{1-D}{LC\cdot s^2+\frac{L}{R_0}\cdot s+(1-D)^2} \tag{93}$$

$$V_o=C(\infty)=V_i\cdot\frac{1}{1-D} \tag{94}$$

**Simulation and physical experiments**

All simulations were conducted using LTspice, where the Boost converter was modeled using time-domain differential equations incorporating parasitic resistances and nonlinear behaviors. The simulation framework enabled the evaluation of system responses under a range of input voltage and load step scenarios. Core components such as inductors, capacitors, and MOSFETs were modeled with realistic ESR and switching characteristics, and a fixed-step solver was used to ensure numerical stability in fast transient simulations.

For experimental validation, a prototype Boost converter circuit was fabricated on a custom-designed, two-layer PCB using standard off-the-shelf components. Careful layout optimization was employed to minimize parasitic wire inductance, resistive losses, and electromagnetic interference. Thermal considerations were addressed through short trace lengths and widened copper planes to improve heat dissipation and reduce resistance drift.

A programmable DC power supply was used to provide a stable input voltage source, allowing for controlled step changes during testing. The power MOSFET was driven by a PWM control signal generated from a calibrated PWM module operating at 10 kHz. Output voltage dynamics were captured using a high-precision digital oscilloscope (SDS2204X, Siglent Technologies), featuring a 2 GSa/s sampling rate and 12-bit vertical resolution.

The steady-state error and the dynamic-state error in this study were calculated as follows:

$$Error = \frac{|y - \hat{y}|}{y} \times 100\%$$

where $y$ represents the value of the sampling point of the actual waveform, and $\hat{y}$ represents the value of the sampling point of the fitted waveform.

The equation used to calculate the root mean squared error (RMSE) is:

$$RMSE = \sqrt{\frac{1}{N}\sum_{i=1}^{N}(y_i - \hat{y}_i)^2}$$

where $y_i$ represents the value of the i-th sampling point of the actual waveform, and $\hat{y}_i$ represents the value of the i-th sampling point of the fitted waveform, and $N$ is the total number of sampling points. A lower RMSE value indicates better fitting accuracy.

# Figures

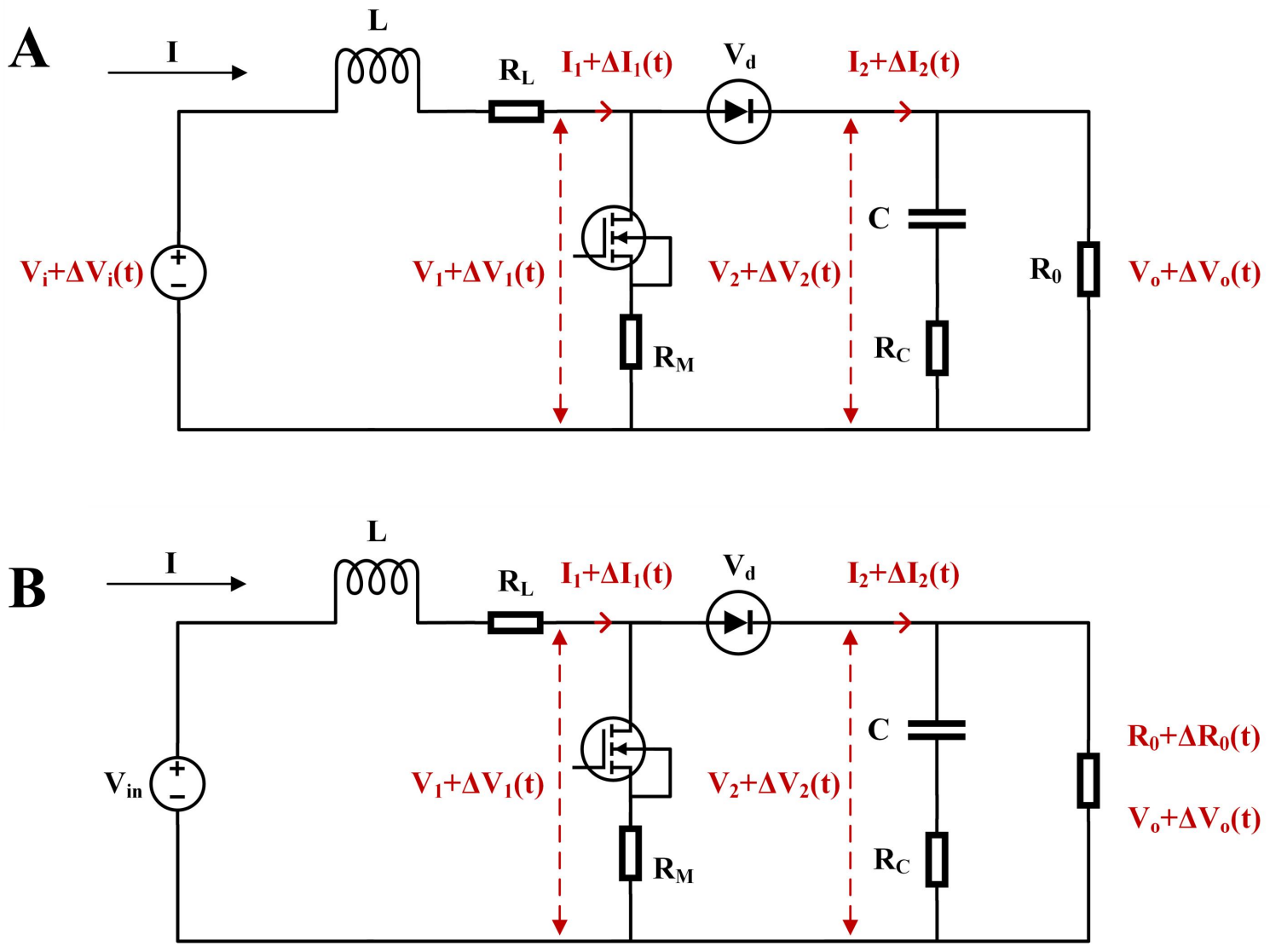

**Fig. S1. Schematic diagrams of the Boost circuit.** By establishing the relationship between the voltages and currents at different nodes before and after the disturbance, the transfer functions of the output voltage with respect to the input voltage (**A**) and load resistance (**B**) can be obtained.

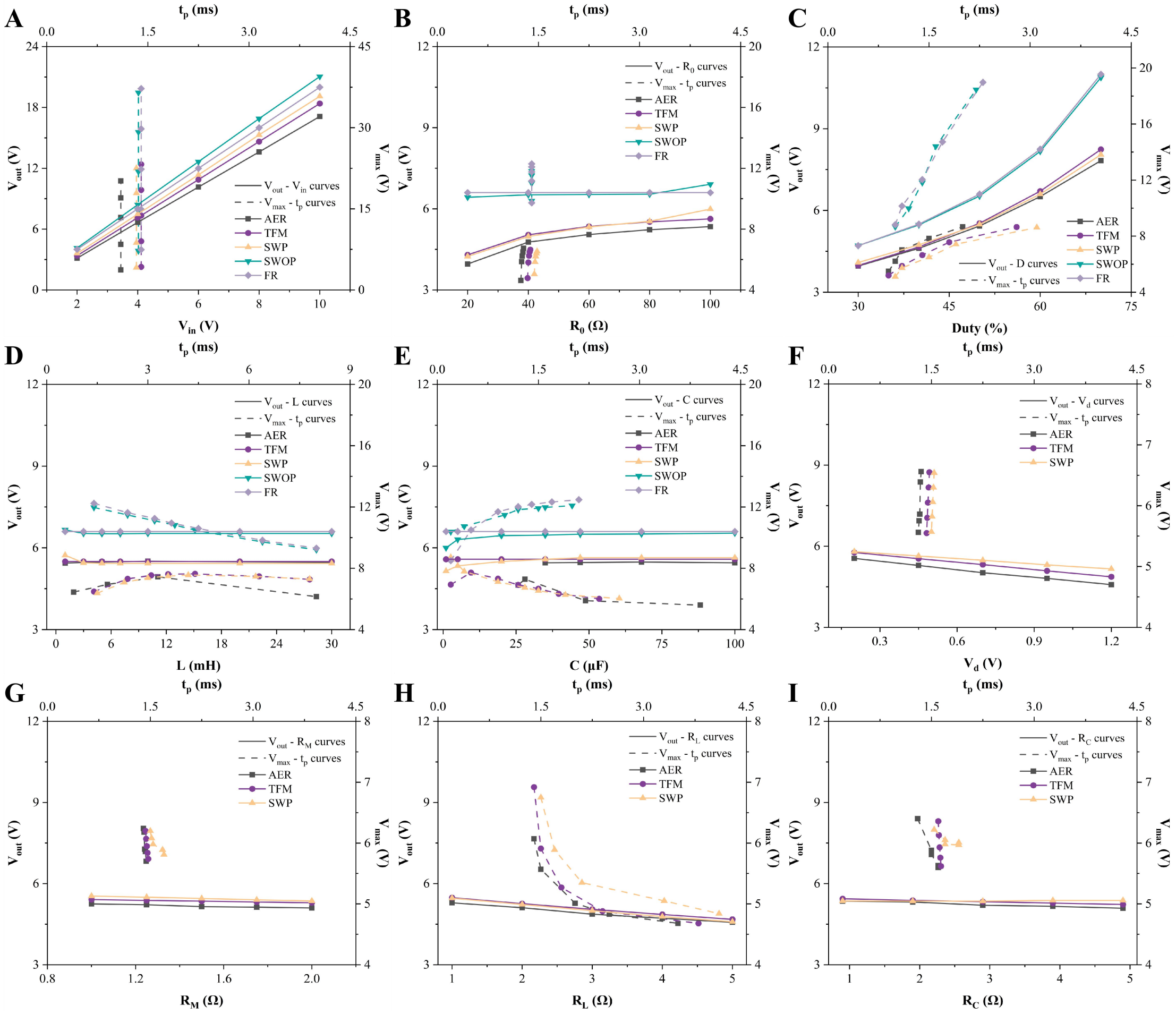

**Fig. S2. Influence of each parameter on output voltages.** (**A**) The solid line represents the $V_{out}$-X curve, and the dashed line represents the $V_{max}$-$t_p$ curve. As the input voltage increases, both $V_{out}$ and $V_{max}$ increase, while $t_p$ remains nearly constant. (**B**) As the load resistance increases, $V_{out}$, $t_p$, and $V_{max}$ all increase. (**C**) As the duty cycle increases, $V_{out}$, $t_p$, and $V_{max}$ all increase. (**D**) As the inductance increases, $V_{out}$ remains nearly constant, $t_p$ increases, and $V_{max}$ initially increases before decreasing. (**E**) As the capacitance increases, $V_{out}$ remains nearly

constant, $t_p$ increases, and $V_{max}$ initially increases before decreasing. (**F**) As the diode forward voltage increases, $V_{out}$, $t_p$, and $V_{max}$ all decrease. (**G**) As the on-resistance of the MOS transistor increases, both $V_{out}$ and $V_{max}$ decrease, while $t_p$ increases. (**H**) As the parasitic resistance of the inductor increases, both $V_{out}$ and $V_{max}$ decrease, while $t_p$ increases. (**I**) As the parasitic resistance of the capacitor increases, both $V_{out}$ and $V_{max}$ decrease, while $t_p$ increases.

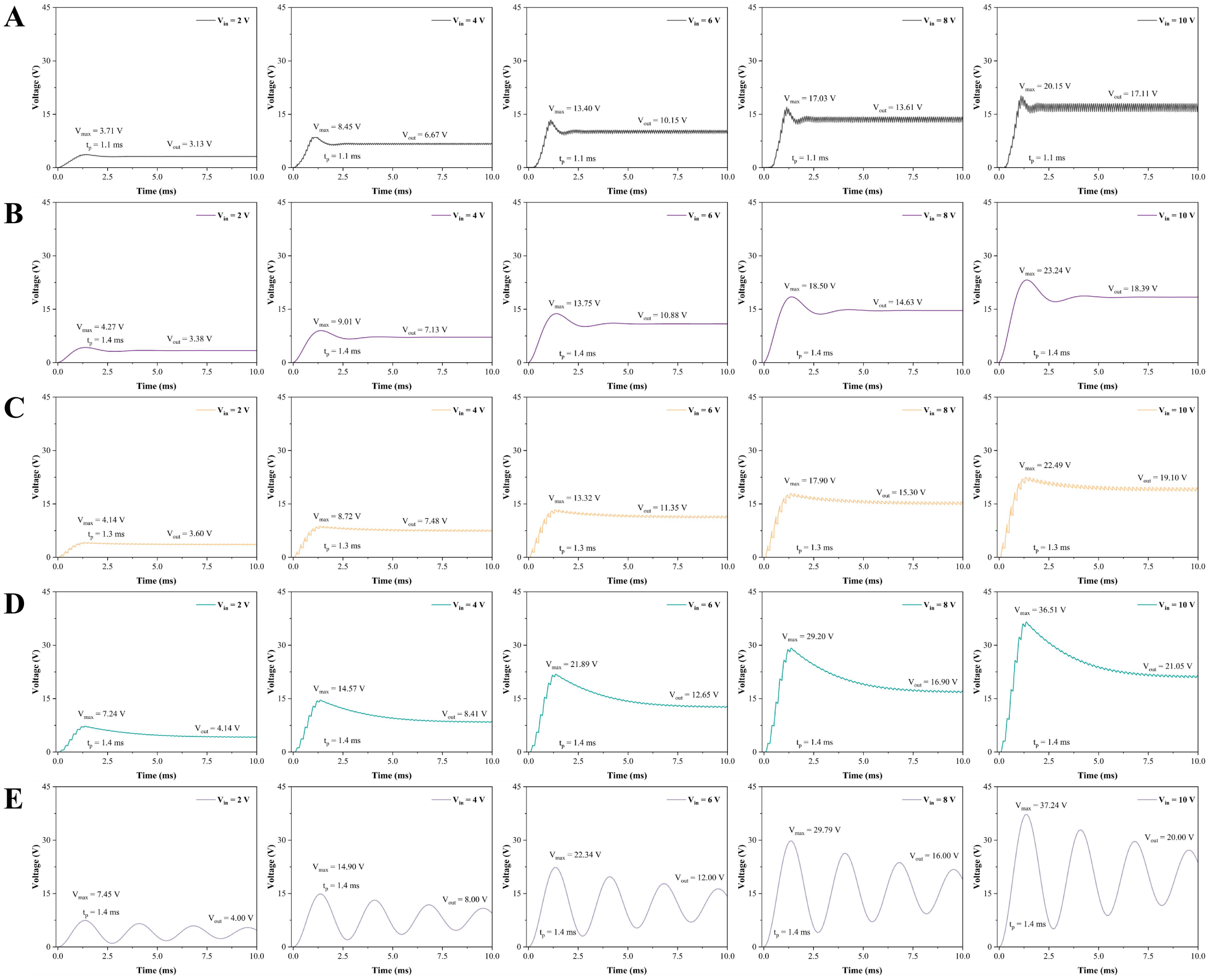

**Fig. S3. Influence of $V_i$ on output voltage.** (**A**) Actual experimental results. (**B**) Transfer function model. The fitting results are in good agreement with the experimental data. (**C**) Simulation with parasitics. While the steady-state performance is satisfactory, there is a partial loss in dynamic performance. (**D**) Simulation without parasitics. Both the steady-state and dynamic performances exhibit significant deviations from the experimental results. (**E**) Formula

from the reference. The oscillation time is excessively long, resulting in a substantial discrepancy with the experimental results.

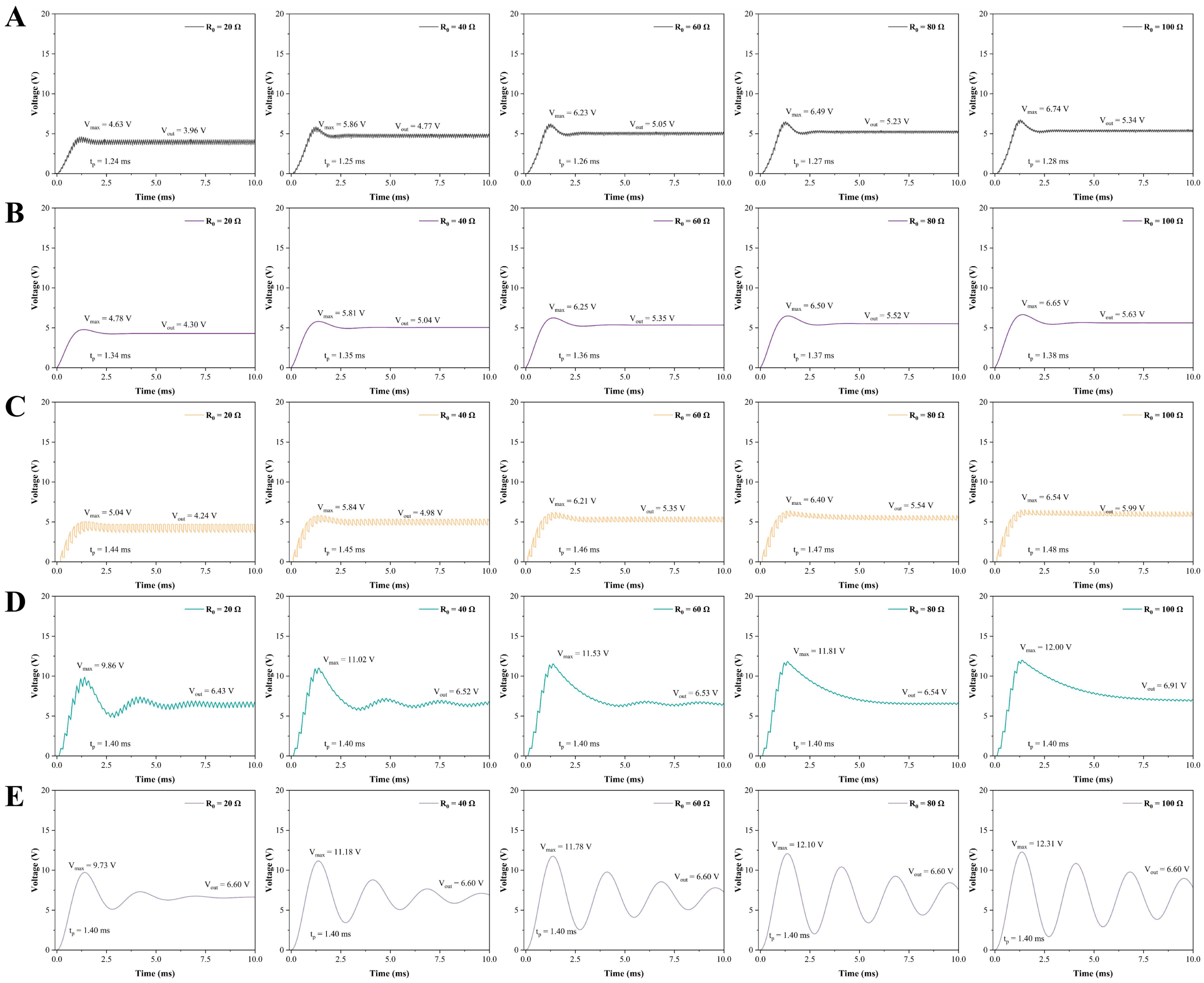

**Fig. S4. Influence of $R_0$ on output voltage.** (**A**) Actual experimental results. (**B**) Transfer function model. The fitting results are in good agreement with the experimental data. (**C**) Simulation with parasitics. While the steady-state performance is satisfactory, there is a partial loss in dynamic performance. (**D**) Simulation without parasitics. Both the steady-state and dynamic performances exhibit significant deviations from the experimental results. (**E**) Formula from the reference. The oscillation time is excessively long, resulting in a substantial discrepancy with the experimental results.

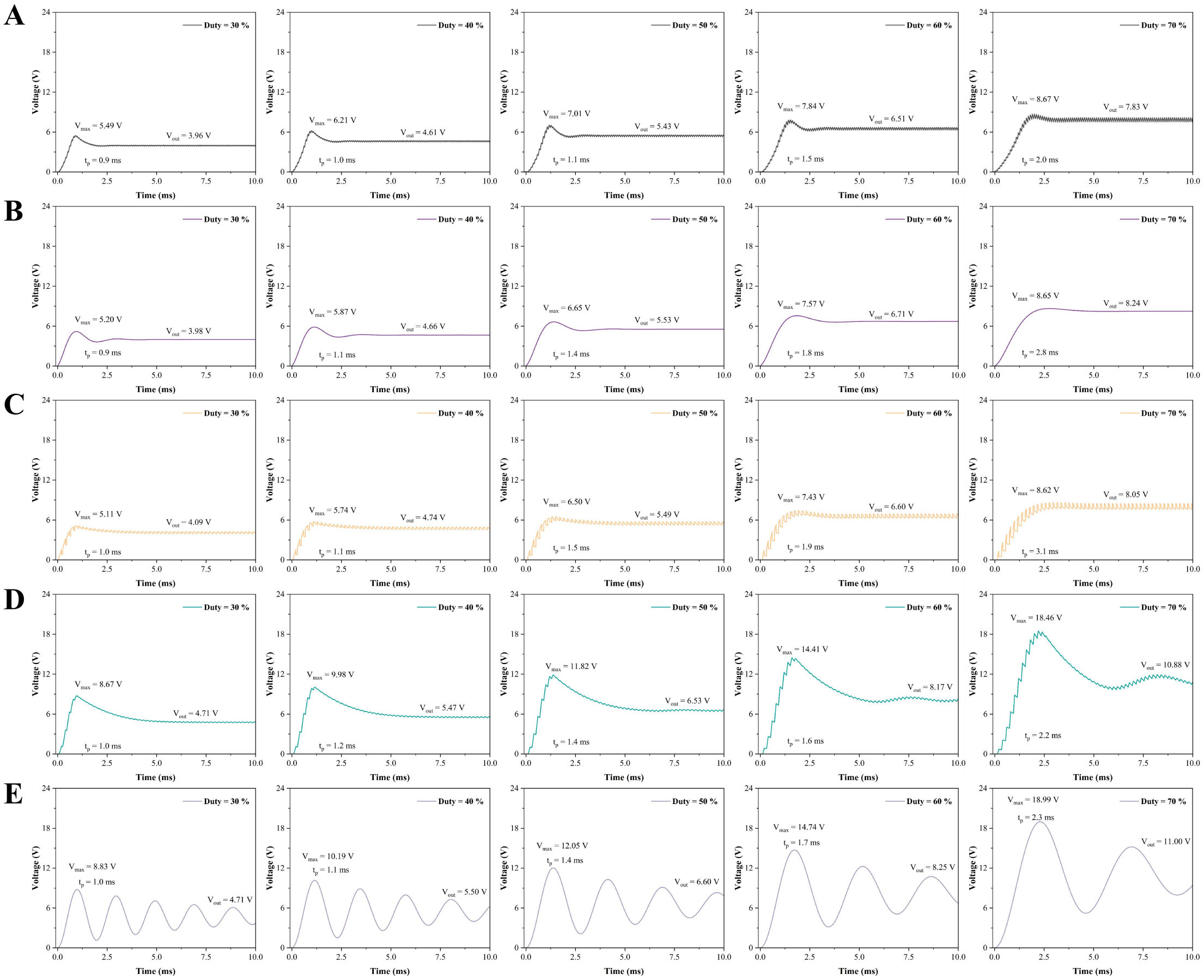

**Fig. S5. Influence of *D* on output voltage.** (**A**) Actual experimental results. (**B**) Transfer function model. The fitting results are in good agreement with the experimental data. (**C**) Simulation with parasitics. While the steady-state performance is satisfactory, there is a partial loss in dynamic performance. (**D**) Simulation without parasitics. Both the steady-state and dynamic performances exhibit significant deviations from the experimental results. (**E**) Formula from the reference. The oscillation time is excessively long, resulting in a substantial discrepancy with the experimental results.

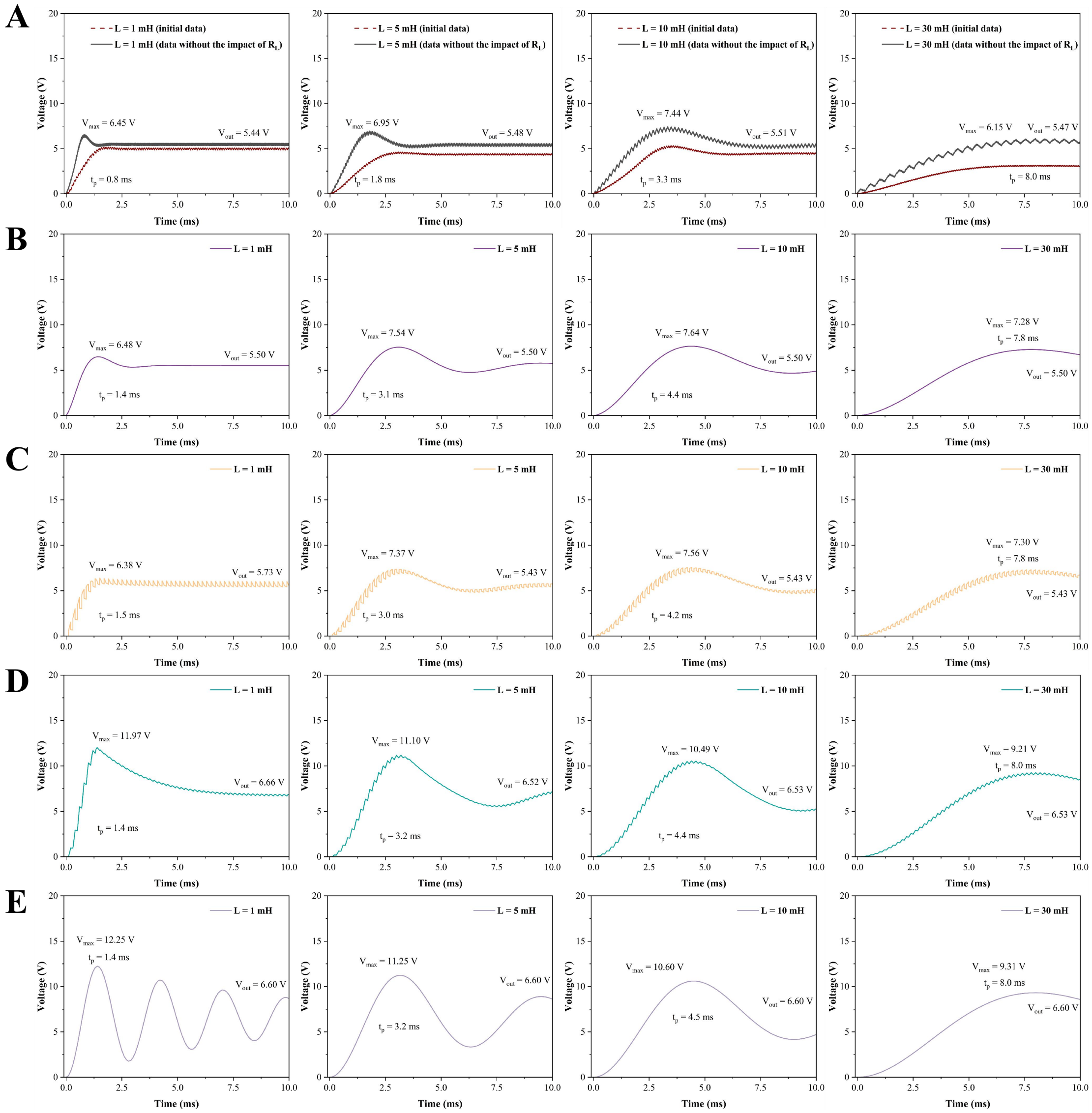

**Fig. S6. Influence of $L$ on output voltage.** (**A**) Actual experimental results (The red line represents the initial data, while the black line represents the data after eliminating the influence of $R_L$ according to the trend shown in fig. S8). (**B**) Transfer function model. The fitting results are in good agreement with the experimental data. (**C**) Simulation with parasitics. While the steady-state performance is satisfactory, there is a partial loss in dynamic performance. (**D**) Simulation without parasitics. Both the steady-state and dynamic performances exhibit significant deviations from the experimental results. (**E**) Formula from the reference. The oscillation time is excessively long, resulting in a substantial discrepancy with the experimental results.

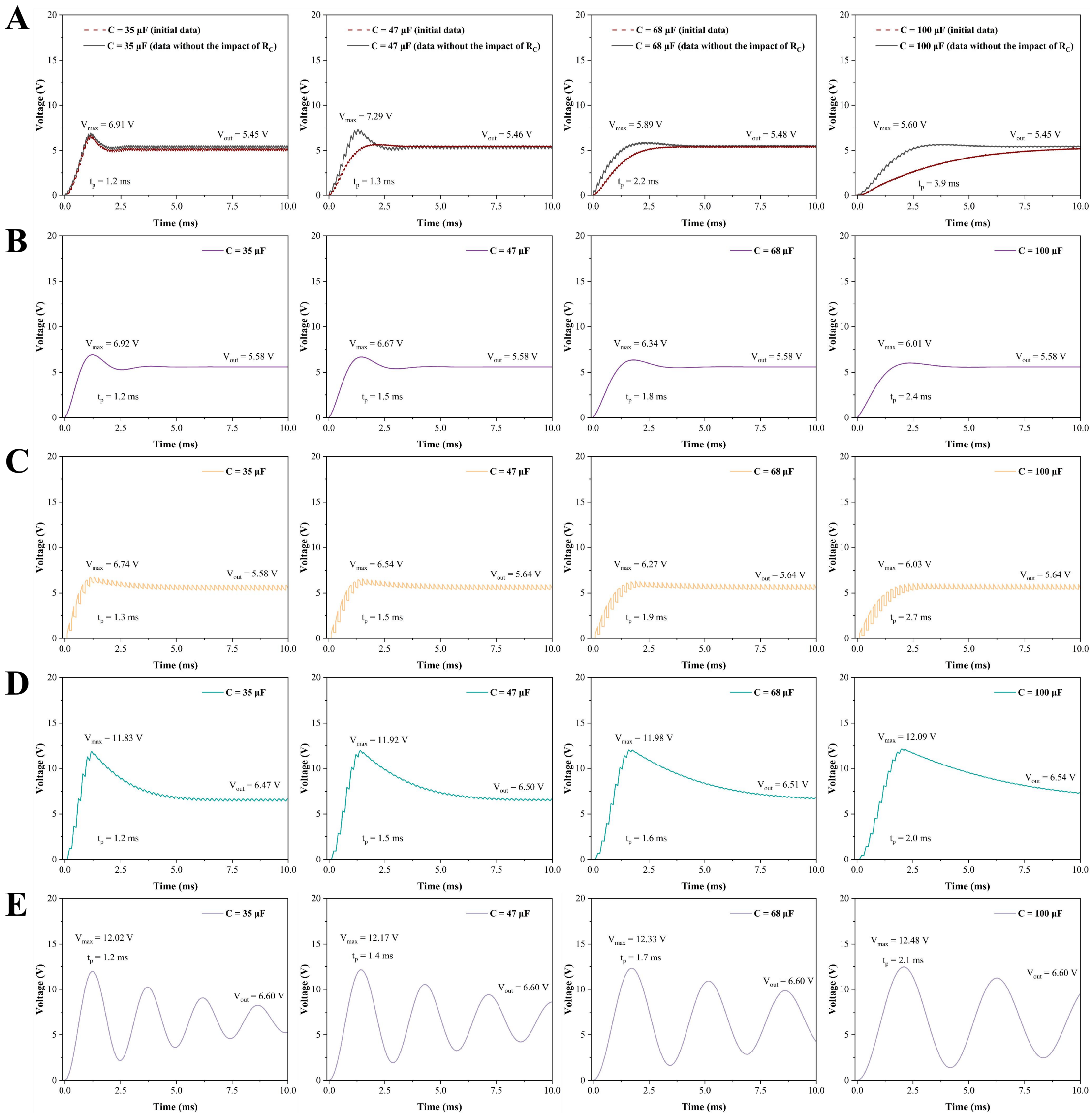

**Fig. S7. Influence of $C$ on output voltage.** (**A**) Actual experimental results (The red line represents the initial data, while the black line represents the data after eliminating the influence of $R_C$ according to the trend shown in fig. S9). (**B**) Transfer function model. The fitting results are in good agreement with the experimental data. (**C**) Simulation with parasitics. While the steady-state performance is satisfactory, there is a partial loss in dynamic performance. (**D**)

Simulation without parasitics. Both the steady-state and dynamic performances exhibit significant deviations from the experimental results. (**E**) Formula from the reference. The oscillation time is excessively long, resulting in a substantial discrepancy with the experimental results.

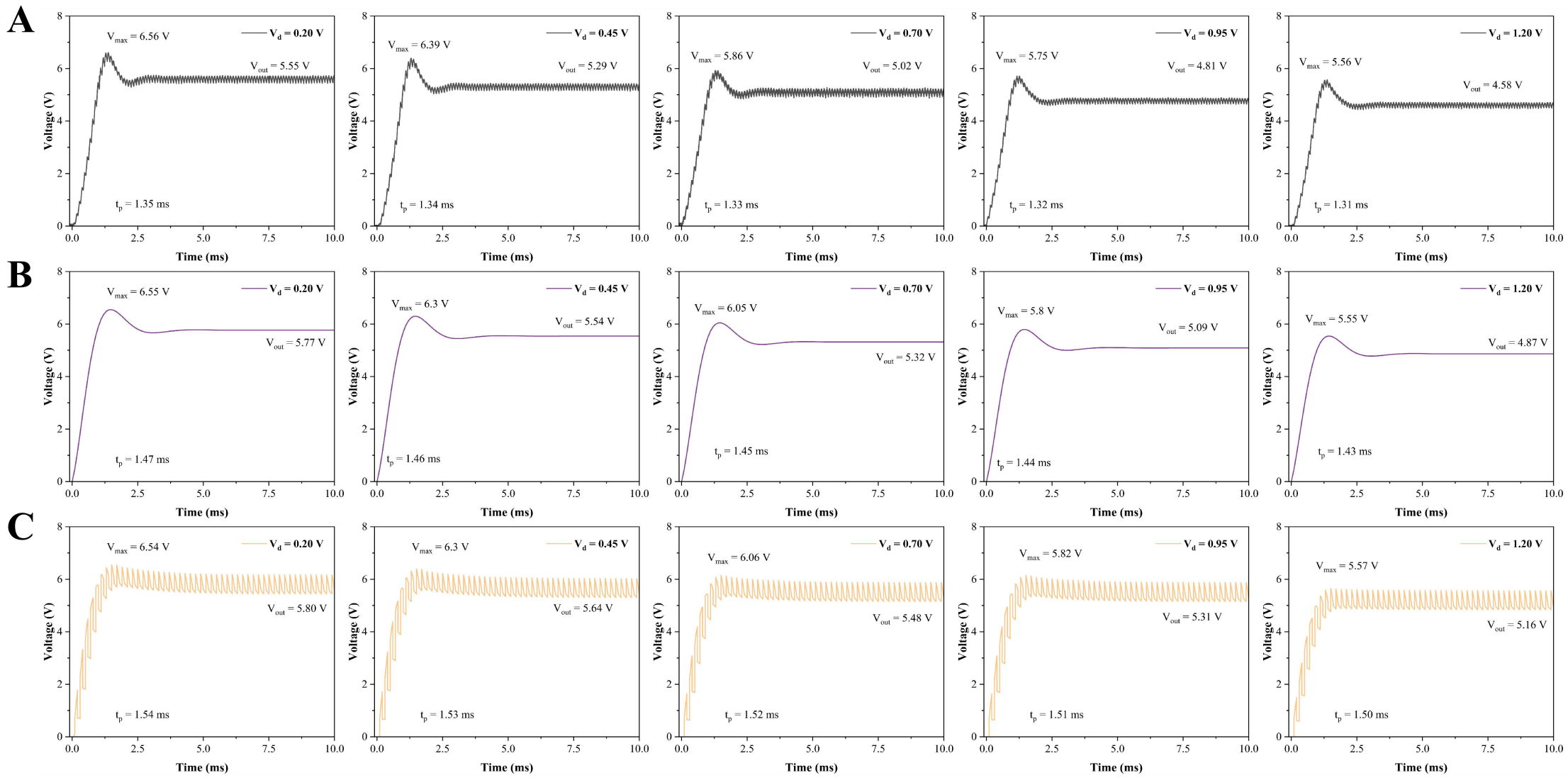

**Fig. S8. Influence of $V_d$ on output voltage.** (**A**) Actual experimental results. (**B**) Transfer function model. The fitting results are in good agreement with the experimental data. (**C**) Simulation with parasitics. While the steady-state performance is satisfactory, there is a partial loss in dynamic performance.

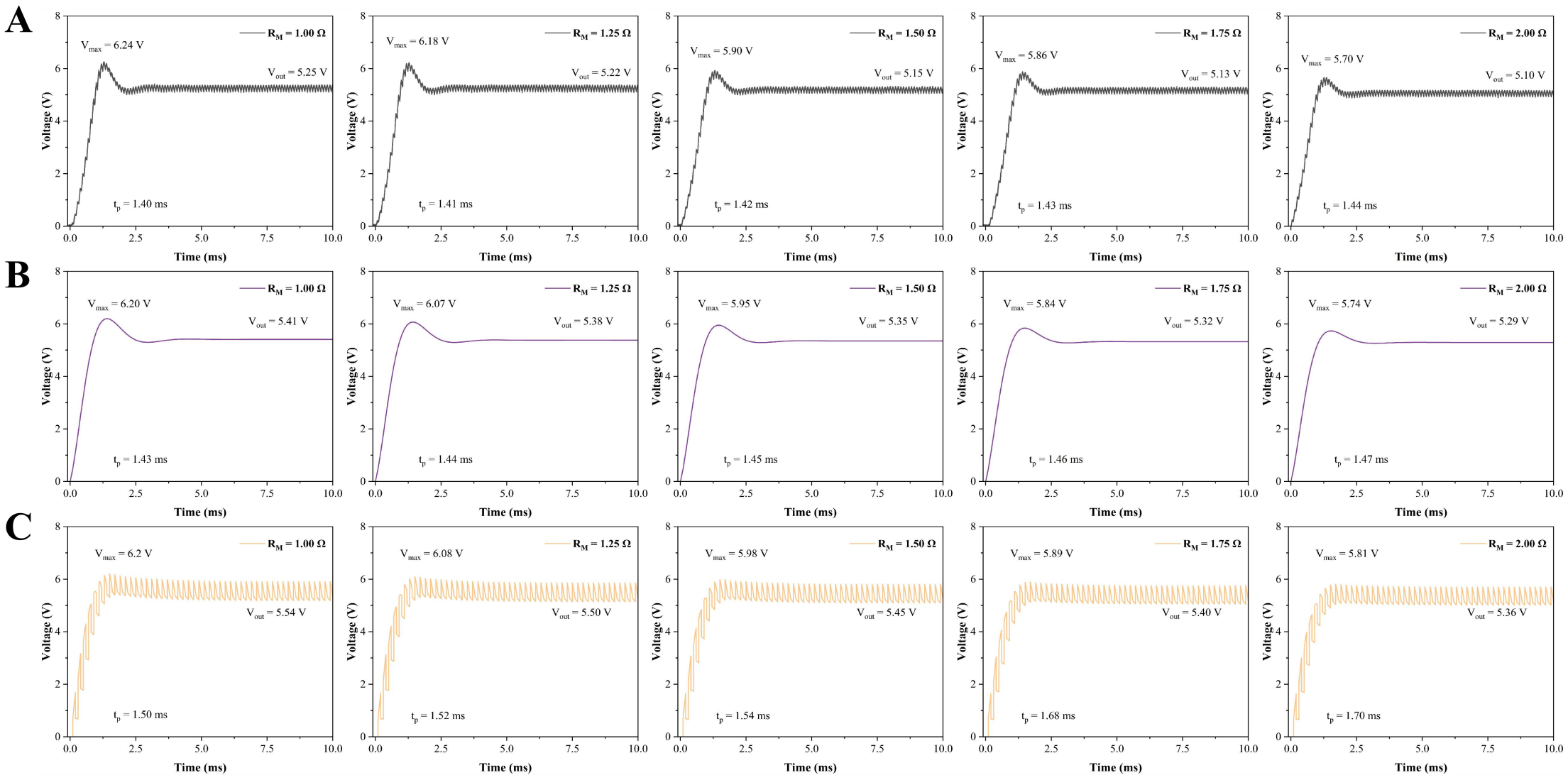

**Fig. S9. Influence of $R_M$ on output voltage.** (**A**) Actual experimental results. (**B**) Transfer function model. The fitting results are in good agreement with the experimental data. (**C**) Simulation with parasitics. While the steady-state performance is satisfactory, there is a partial loss in dynamic performance.

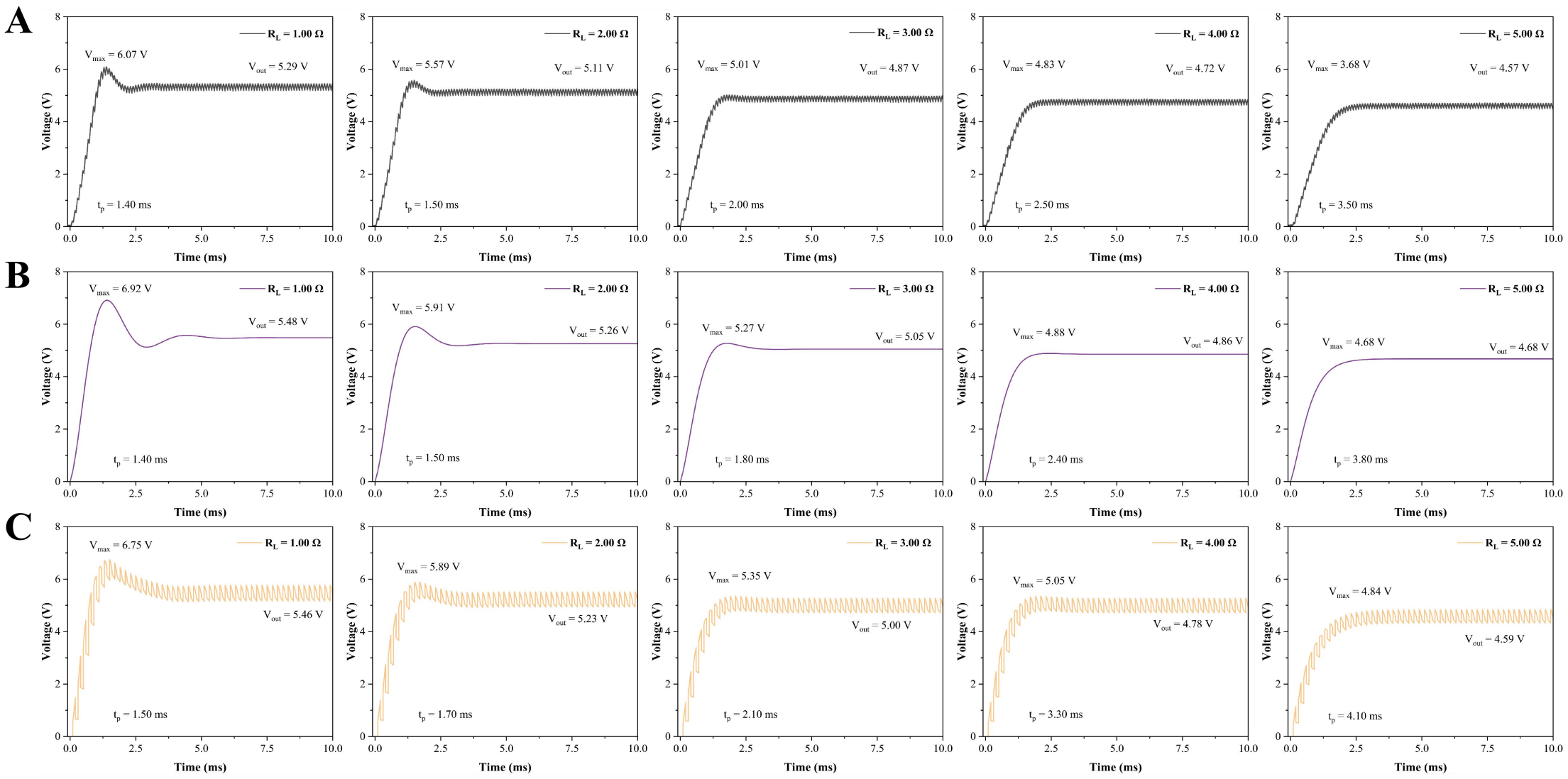

**Fig. S10. Influence of $R_L$ on output voltage.** (**A**) Actual experimental results. (**B**) Transfer function model. The fitting results are in good agreement with the experimental data. (**C**)

Simulation with parasitics. While the steady-state performance is satisfactory, there is a partial loss in dynamic performance.

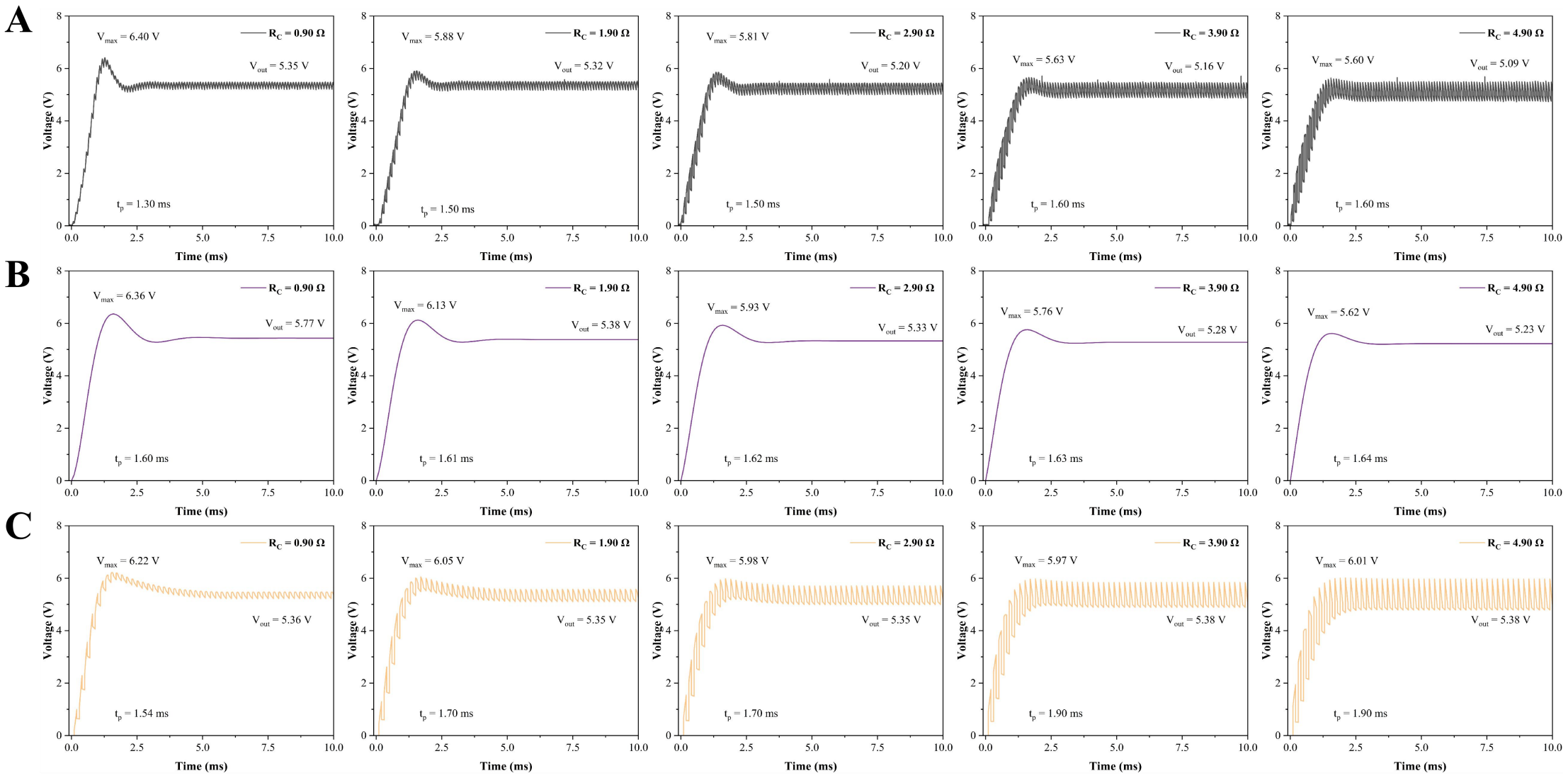

**Fig. S11. Influence of $R_C$ on output voltage.** (**A**) Actual experimental results. (**B**) Transfer function model. The fitting results are in good agreement with the experimental data. (**C**) Simulation with parasitics. While the steady-state performance is satisfactory, there is a partial loss in dynamic performance.

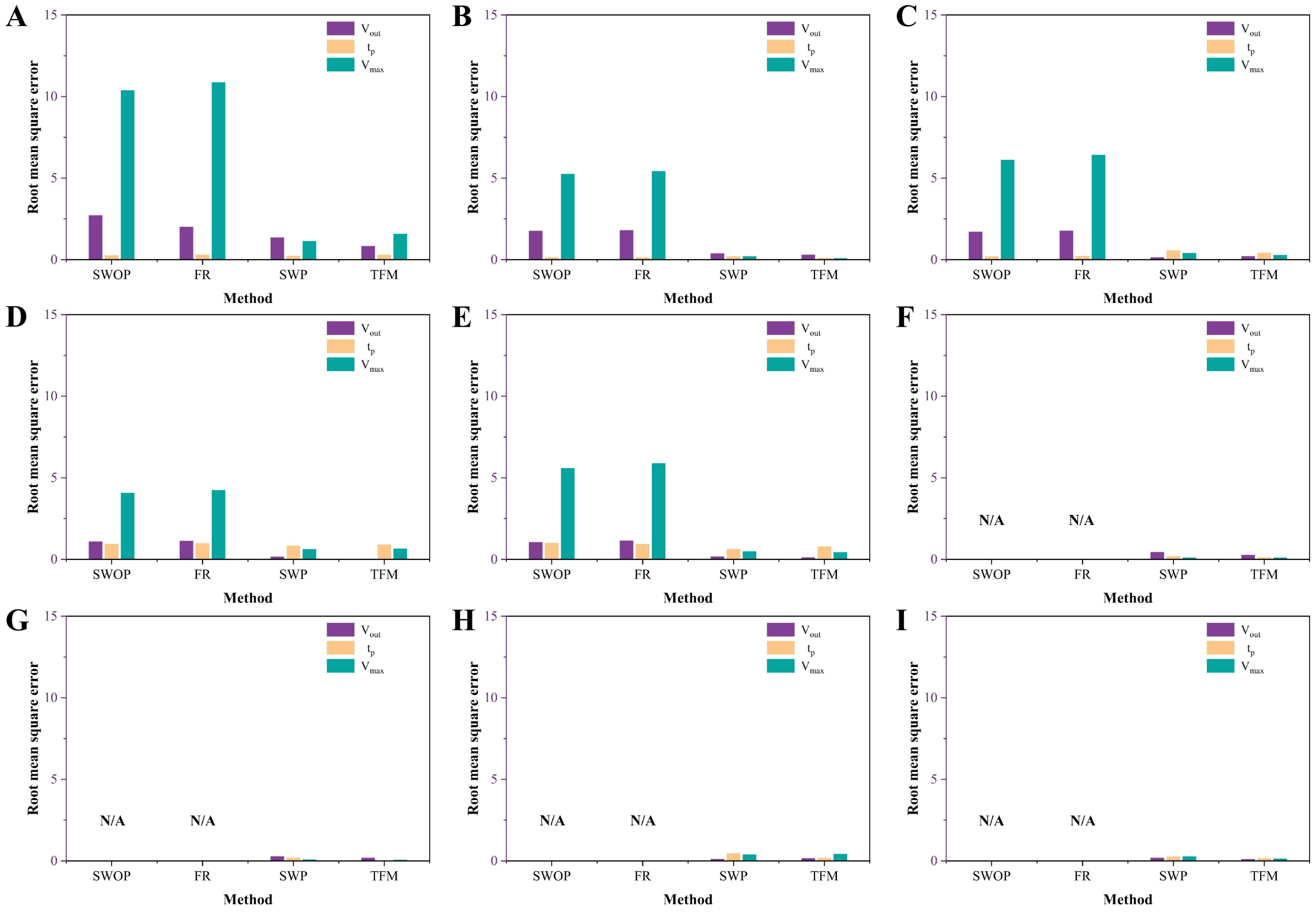

**Fig. S12. Analysis of fitting errors for different models.** (**A** to **E**) As $V_i$, $R_0$, $D$, $L$, and $C$ vary, the root mean square errors of $V_{out}$ and $V_{max}$ for TFM are significantly reduced. (**F** to **I**) As $V_d$, $R_M$, $R_L$, and $R_C$ vary, the root mean square errors of $V_{out}$ and $V_{max}$ for TFM remain similarly small.

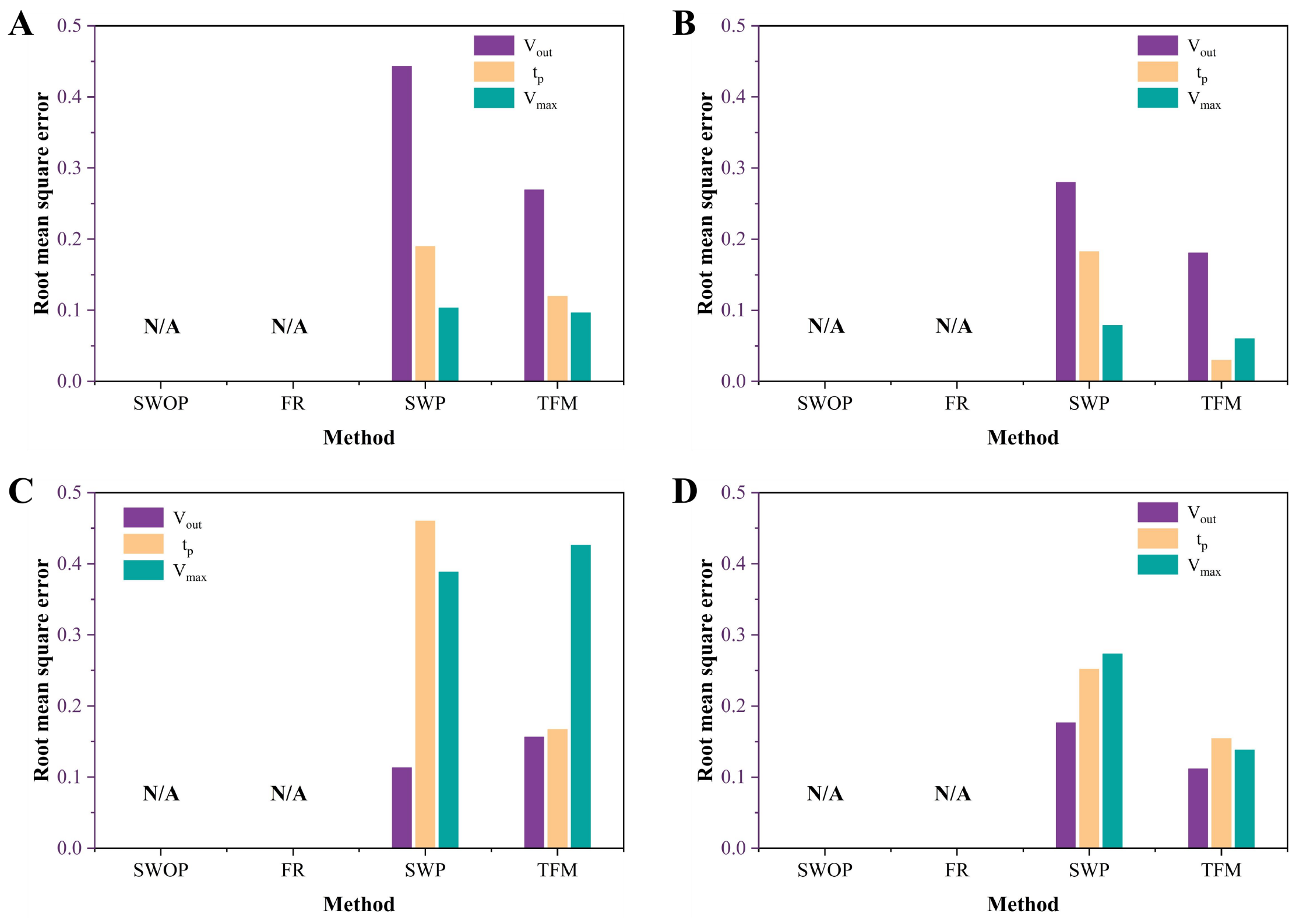

**Fig. S13. Magnified plot of the RMSE.** (**A** to **D**) As $V_d$, $R_M$, $R_L$, and $R_C$ vary, all root mean square errors for TFM remain below 0.5.

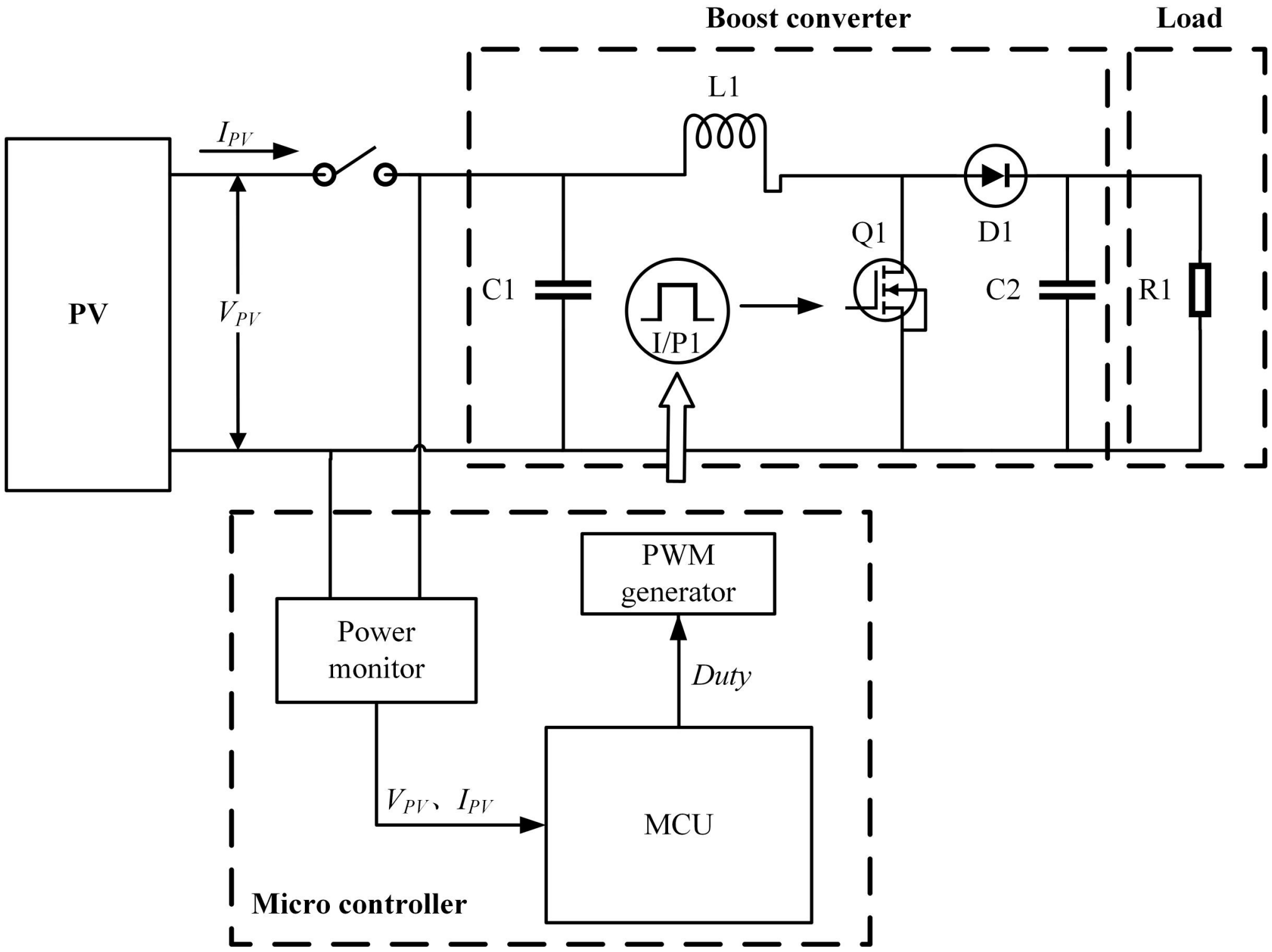

**Fig. S14. Schematic diagram of the spaceborne Boost system.** In validating the performance of the Boost converter, the input voltage is varied by adjusting the number of connected photovoltaic panels, while the load resistance is modified by changing the operating mode of the load (an optical micro-camera). The duty cycle is maintained at a fixed value throughout.

## Tables

**Table S1**. Parameter values of different methods (when $V_i$ changes).

| | Nominal value | Method 1 | Method 2 | Method 3 |
|---|---|---|---|---|
| $V_i$ | 3.3 V | 3.3 V | 3.3 V | 3.3 V |
| $L$ | 1±0.1 mH | 1 mH | 1 mH | 20 mH |
| $R_L$ | 1.4±0.1 Ω | 1.5 Ω | 1.4 Ω | 1.4 Ω |
| $C$ | 47±5 μF | 42 μF | 42 μF | 5 μF |
| $R_C$ | 1.2±0.1 Ω | 1.3 Ω | 1 Ω | 1 Ω |
| $R_M$ | 0.8±0.1 Ω | 0.9 Ω | 0.8 Ω | 0.8 Ω |
| $V_d$ | 0.4±0.1 V | 0.5 V | 0.4 V | 0.4 V |
| $R_0$ | 95±5 Ω | 92 Ω | 95 Ω | 95 Ω |
| $D$ | 0.50±0.02 | 0.49 | 0.50 | 0.40 |

**Table S2**. Error analysis of different models (when $V_i$ changes).

| | $V_o$ (V) | Steady-state error (%) | $V_{max}$ (V) | Dynamic-state error (%) |
|---|---|---|---|---|
| Actual experimental results | 5.35 | - | 6.74 | - |
| $V_o = \frac{V_i}{1-D}$ | 6.47 | 20.9 | - | - |
| Formula from the reference | 6.47 | 20.9 | 11.94 | 77.1 |
| Energy-based model | 5.45 | 1.9 | 6.71 | 0.4 |
| Transfer function model | 5.45 | 1.9 | 6.40 | 5.0 |

**Table S3**. Parameter values of different methods (when $R_0$ changes).

| | Nominal value | Method 1 | Method 2 | Method 3 |
|---|---|---|---|---|
| $V_i$ | 5.0 V | 5.0 V | 5.0 V | 5.0 V |
| $L$ | 1±0.1 mH | 1 mH | 1 mH | 1 mH |
| $R_L$ | 1.4±0.1 Ω | 1.4 Ω | 1.4 Ω | - |
| $C$ | 47±5 μF | 43 μF | 47 μF | 47 μF |
| $R_C$ | 1.1±0.1 Ω | 1.0 Ω | 1.2 Ω | - |
| $R_M$ | 0.8±0.1 Ω | 0.8 Ω | 0.8 Ω | - |
| $V_d$ | 0.4±0.1 V | 0.4 V | 0.4 V | - |
| $R_0$ | 10±2 Ω | 10 Ω | 8.5 Ω | 8.5 Ω |
| $\Delta R_0$ | 140±4 Ω | 140 | 141.5 | 141.5 |
| $D$ | 0.50±0.02 | 0.50 | 0.50 | 0.50 |

**Table S4**. Error analysis of different models (when $R_0$ changes).

| | $V_o$ (V) | Steady-state error (%) | $V_{max}$ (V) | Dynamic-state error (%) |
|---|---|---|---|---|
| Actual experimental results | 8.80 | - | 13.43 | - |
| Simulation without parasitics | 10.15 | 15.3 | 19.09 | 42.1 |
| Energy-based model | 9.10 | 3.4 | 12.56 | 6.5 |
| Transfer function model | 8.67 | 1.5 | 13.27 | 1.2 |

**Table S5**. Root mean squared error for different parameter changes.

| | RMSE of $V_o$ | | | | RMSE of $V_{max}$ | | | |
|---|---|---|---|---|---|---|---|---|
| | SWOP | FR | SWP | TFM | SWOP | FR | SWP | TFM |
| $V_i$ | 2.71 | 2.00 | 1.35 | 0.83 | 10.37 | 10.87 | 1.14 | 1.58 |
| $R_0$ | 1.76 | 1.80 | 0.38 | 0.30 | 5.25 | 5.42 | 0.21 | 0.08 |
| $D$ | 1.71 | 1.78 | 0.14 | 0.21 | 6.11 | 6.41 | 0.40 | 0.28 |
| $L$ | 1.09 | 1.13 | 0.15 | 0.04 | 4.07 | 4.24 | 0.62 | 0.65 |
| $C$ | 1.05 | 1.14 | 0.17 | 0.12 | 5.59 | 5.89 | 0.48 | 0.43 |
| $V_d$ | - | - | 0.44 | 0.27 | - | - | 0.10 | 0.10 |
| $R_M$ | - | - | 0.28 | 0.18 | - | - | 0.08 | 0.06 |
| $R_L$ | - | - | 0.11 | 0.16 | - | - | 0.39 | 0.43 |
| $R_C$ | - | - | 0.18 | 0.11 | - | - | 0.27 | 0.14 |

**Table S6**. Error analysis of output voltage waveform predictions (TFM).

| | $V_o$ (V) | Steady-state error (%) | $V_{max}$ (V) | Dynamic-state error (%) |
|---|---|---|---|---|
| Initial waveform | 8.80 | 1.5 | 13.43 | 1.2 |
| **Fig. 5A** | 8.87 | 3.3 | 9.35 | 1.0 |
| **Fig. 5B** | 8.79 | 3.8 | 9.09 | 6.9 |
| **Fig. 5C** | 8.11 | 2.8 | 9.10 | 0.2 |
| **Fig. 5D** | 8.79 | 4.0 | 8.84 | 2.3 |
| **Fig. 5E** | 8.74 | 5.7 | 9.66 | 0.8 |
| **Fig. 5F** | 8.60 | 2.1 | 9.17 | 1.5 |
| **Fig. 5G** | 8.34 | 1.7 | 8.40 | 3.3 |
| **Fig. 5H** | 8.40 | 2.6 | 9.04 | 9.5 |